\documentclass[smallextended]{svjour3}

\usepackage{amsfonts,amssymb}
\usepackage[fleqn]{amsmath}
\usepackage{amssymb}
\usepackage{dsfont} 
\usepackage{bbm,bm} 
\usepackage{psfrag}
\usepackage{graphicx}     
\usepackage{slashed}  
\usepackage{color}
\usepackage{float}
\usepackage{multirow}

\allowdisplaybreaks

\newcommand{\en}{\enspace}
\newcommand{\Id}{\mathds{1}}

\newcommand{\tn}[1]{\textnormal{#1}}
\newcommand{\bra}[1]{\left\langle #1 \right|}
\newcommand{\ket}[1]{\left| #1 \right\rangle}

\newcommand{\produkt}[2]{#1_{\mu} #2^{\mu}}

\newcommand{\ga}{\gamma}

\newcommand{\sla}[1]{\slashed{#1}}

\newcommand{\citeexplicit}[2]{\cite{#1}} 
\newcommand{\baryongeneral}[1]{H_b^{j= #1}} 
\newcommand{\baryongeneralW}{H_b^{j}} 
\newcommand{\vectorspinor}{R} 
\newcommand{\nonlocaloperator}{O}
\newcommand{\trace}[1]{\tn{Tr}\left[{#1}\right]}

\def\makeheadbox{{%
\hfill
$\textnormal{{\small\textsc DESY 12-225}}$ \newline
\phantom{a}\hfill
$\textnormal{{\small\textsc DO-TH 12/37}}$\\
}%
\hss}

\begin{document}
 
\title{
Light-Cone Distribution Amplitudes of the Ground State Bottom Baryons in HQET
}

\author{A.~Ali, 
        C.~Hambrock,
        A.\,Ya.~Parkhomenko,
        Wei~Wang}

\institute{A.~Ali \at
              \small{\em Deutsches Elektronen-Syn\-chro\-tron DESY} \\ 
              \email{ahmed.ali@desy.de}          
           \and
           C.~Hambrock \at
              \small{\em Technische Universit\"at Dortmund} \\ 
              \email{christian.hambrock@tu-dortmund.de}  
           \and
           A.\,Ya.~Parkhomenko \at
               \small{\em P.\,G.~Demidov Yaroslavl State University} \\ 
               \email{parkh@uniyar.ac.ru}  
           \and
           Wei Wang \at
               \small{\em Deutsches Elektronen-Syn\-chro\-tron DESY}\\ 
               \email{wei.wang@desy.de}  
                    \at 
               \small{Address after October 1, 2012:  
                      \em Helmholtz-Institut f\"ur Strahlen und Kernphysik, 
                          Universit\"at Bonn, Bonn 53115, Germany} \\ 
               \email{weiwang@hiskp.uni-bonn.de}  
}

\date{Received: date / Accepted: date}

\date{\today}

\maketitle

\begin{abstract} 
We provide the definition of the complete set 
of light-cone distribution amplitudes (LCDAs) 
for the ground state heavy bottom baryons with 
the spin-parities $J^P = 1/2^+$ and $J^P = 3/2^+$  
in the heavy quark limit. 
We present the renormalization effects on the twist-2 
light-cone distribution amplitudes and  use the 
QCD sum rules to compute the moments of twist-2, 
twist-3, and twist-4 LCDAs. 
Simple models for the heavy baryon distribution 
amplitudes are analyzed with account of their 
scale dependence. 
\end{abstract}

\maketitle

\section{Introduction}
Precision tests of the unitarity of the quark mixing matrix remain high on the agenda of
flavor physics. They allow to pin down the
 Standard Model~(SM) description of CP-violation and may reveal physics beyond the SM (BSM).  
On the experimental side, the two $B$-meson factories at SLAC and KEK, after 
approximately a decade of their operation,  have made a great impact on the origin of CP-violation
 in the quark sector of the SM. 
Especially in $b$ physics, the $B$-factory experiments
BABAR and BELLE have concentrated on the production and decays of the $b \bar{b}$ resonances,
 $\Upsilon(4S)$ and $\Upsilon(5S)$, yielding precise measurements of their decay products,
the  $B$- and the $\bar{B}$-mesons.
Theoretically, inclusive decays of the $B$-mesons, such as the radiative and semileptonic decays,
are under quantitative control, thanks to  the use of the Heavy Quark Effective Theory (HQET), allowing an
expansion in the inverse $b$-quark mass, with perturbative QCD allowing to calculate corrections in
$\alpha_s(m_b)$ to each order in $1/m_b$. Exclusive semileptonic and radiative decays require, on the other
hand, a precise knowledge of the non-perturbative quantities - the decay form factors - to be analyzed
precisely. However, also in such cases,
the large energy released in the $B \to h X$ decays, where $h$ is a light meson and
 $X=\gamma, \ell^+\ell^-, \ell^\pm \nu_\ell$,
 allows to relate several of these form factors. The resulting large energy
 effective theory (LEET), which has been replaced by a more systematic effective theory, the Soft Collinear Effective
 Theory (SCET),
has been put to good use in these decays, resulting in  a number of theoretical predictions on different observables
 in various channels, which are  found in global agreement with the experimental measurements
 (see Ref.~\cite{Buchalla:2008jp} for a review).

The attention in flavor physics has so far mostly been focused on the meson sector.
 Specific  processes involving bottom baryons, such as rare decays 
involving flavor-changing neutral current (FCNC) 
transitions, are also potential sources of  BSM physics.
Experiments at the LHC, in particular the
 LHC$b$, are already analyzing the copiously produced $b$-baryons. As more luminosity is collected,
the $b$-baryon sector will become a wider and quantitative field. Theoretically, these decays are more involved,
in particular the non-leptonic $b$-baryon decays, but the semileptonic and radiative $b$-baryon decays are 
tractable. As the starting point for the calculations of the transition form factors, a
precise knowledge of  the quark distributions inside the baryons  is needed.
 The LCDAs provide this input in the case of light quarks having high virtuality, which then become 
a crucial input to the calculations of the baryonic transitions in the light cone sum rule approach.  
Baryonic transitions involving a heavy quark have been studied since the beginning of the 90s.
 However, the theoretical precision achieved so far is not comparable
to the one attained  for the corresponding mesonic transitions. So far, most  
of the heavy baryon distribution amplitudes discussed in the literature~\cite{Loinaz:1995wz,Hussain:1990uu} 
are the ones which are motivated by various quark models, which are not manifestly  
consistent with the QCD constraints. Hence, it is desirable to make theoretical inroads in this sector, guided
by QCD. This effort will pay off, as the   baryons have
the advantage over the $B$-mesons of providing access to spin correlations in various baryon-to-baryon transitions. 
A full angular analysis of the semileptonic decays of some $b$-baryons,
such as the decays of the $\Lambda_b$ and the $\Omega_b$, will provide a much more detailed view of the underlying dynamics.
Of special interest are the semileptonic heavy baryon decays, governed  by heavy-to-light form factors.  In this retrospect, 
apart from the theoretical analysis  based on the heavy quark effective
 theory~\cite{Isgur:1990pm,Mannel:1990vg,hep-ph/9701399},
the simplification of   baryonic form factors in the large recoil limit is  exploited  in Refs.~\cite{Feldmann:2011xf,Mannel:2011xg}
(see Ref.~\cite{Hiller:2001zj,hep-ph/0702191,Wei:2009np} for an earlier discussion).   In the framework of the soft-collinear
effective theory,  it is demonstrated that  the leading-power heavy-to-light baryonic form factors  at large recoil  obey
the heavy quark and large energy symmetries~\cite{Wang:2011uv}, and in particular the universal  function (soft form factor)
in $\Lambda_b\to \Lambda$ is also calculated within the light-cone QCD sum rules in conjunction with the effective field
theory~\cite{Feldmann:2011xf}. 
\begin{figure}[t]
\centering
\vspace{-0.1cm}
\includegraphics[width=0.6\textwidth]{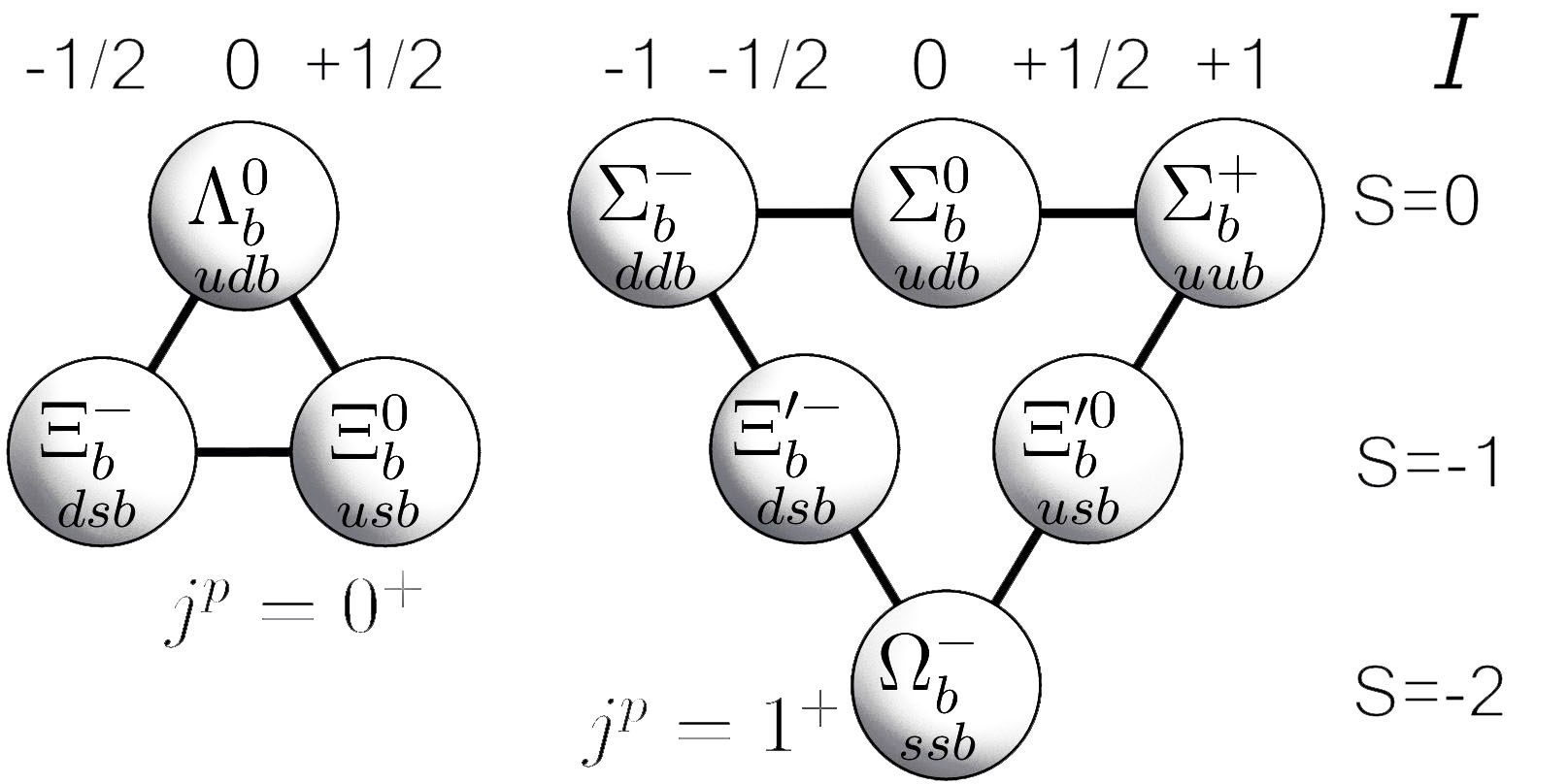}
\caption{
The $SU(3)_{\rm F}$ flavor multiplets of the ground-state bottom baryons.
The flavor $SU(3)_{\rm F}$ triplet with the spin-parity $J^P = 1/2^+$ 
and the scalar light-quark current ($j^p = 0^+$) is shown to the left,  
and the sextet with the spin-parity $J^P = 1/2^+$ and the vector 
light-quark current ($j^p = 1^+$) is shown to the right 
(the similar sextet with $J^P = 3/2^+$ is not presented explicitly).  
The isospin~$I$ and the strangeness~$S$ of each baryon state
are explicitly specified.  
\index{Multiplets of $b$-baryon ground states}
}
\vspace{-0.2cm}
\label{fig:SU(3)-multiplets}
\end{figure}
 
In a pioneering paper~\cite{Ball:2008fw}, the complete 
classification of three-quark light-cone distribution 
amplitudes (LCDAs) of the $\Lambda_b$-baryon 
in QCD in the heavy quark limit was given and 
the scale dependence of the leading-twist LCDA was
discussed. In addition, simple models of the LCDAs 
were suggested and their parameters were fixed 
based on estimates of the first few moments by 
the QCD sum rules method. This analysis can be extended to all the ground-state $b$-baryons with
both the spin parities $J^P=1/2^{+}$ and $J^P=3/2^{+}$. The basic steps and some of the results of
such an analysis were presented in a conference proceedings~\cite{Ali:2012zza}.
 In this paper, we provide the details of the calculations, in particular, concerning the
$J^P=3/2^{+}$ $b$-baryons.
Our generalization of the work~\cite{Ball:2008fw}  to the full ground state $b$-baryon multiplets 
 provides new input for
the eventual calculation of the transition form factors in their decays. In addition, the calculated $SU(3)_{F}$ breaking effects
presented here are of interest for studying the strange-quark properties in hadronic background in general.
Especially the $\Xi_b$-resonances include three different mass quarks, the bottom, the strange and one light $u$- or $d$-quark.
They provide interesting insights on the nature of the strange-quark propagating in the QCD background and give some hints
on the validity of the quark condensate models. 
Comparison with the $K$-mesons  and the  $B_s$-meson will show to what extent the condensates are universal quantities.

This paper is organized as follows: We start by
discussing the  general properties of the single-heavy baryons in the heavy quark limit in
 Sec.~\ref{sec_bprophqlim},
 introduce the local interpolating operators of heavy baryons in HQET in
 Sec.~\ref{sec_locop}, and discuss 
the non-local light-cone operators of these baryons in Sec.~\ref{sec_nonlocop}.  
Section~\ref{sec_renorm} is about the renormalization and the scale-dependence of the various matrix elements
introduced earlier. 
The calculations of the correlation functions, which are defined as matrix elements of the non-local and local operators
are performed subsequently in Sec.~\ref{sec_corrfun} followed by the discussion of our non-perturbative model
 in Sec.\ref{lcdamodel}. Sec.~\ref{sec_resultsLCDAs} contains our results and numerical analysis, and  we conclude with
 a summary in Sec.~\ref{ch_can and out}.

\section{Baryon properties in the HQ limit} 
\label{sec_bprophqlim}

Baryons with one heavy quark~$Q = c, b$ in HQET 
are classified according to the angular momentum~$\ell$
and parity~$p$ of the light quark pair, called diquark. 
The heavy quarks are non-relativistic particles which 
decouple from the diquark in the leading order of the 
$1/m_Q$ expansion. 
The ground-state baryons ($\ell = 0$) with spin-parity~$J^P$ 
are characterized by the spin-parity~$j^p$ of the diquark.
The spins of the light quarks produce two states 
with $j^p = 0^+$ and $j^p = 1^+$. The spin wave-function 
is antisymmetric for the state with $j^p = 0^+$,
while Fermi statistics of the baryon state and antisymmetry
in color space require an antisymmetric flavor wave-function.
This results in a baryonic state with isospin $I = 0$
constructed from the light~$u$- and $d$-quarks and the 
heavy quark~$Q$ which is called the $\Lambda_Q$-baryon 
(the spin-parity is $J^P = 1/2^+$).
When the spin-parity of the diquark is $j^p = 1^+$, 
the spin part of the baryon wave-function is symmetric which 
requires symmetry of the wave-function in the flavor space.
In the case of light~$u$- and $d$-quarks beside the heavy 
quark~$Q$, this gives rise to two degenerate states with 
the isospin $I = 1$, which are called~$\Sigma_Q$- and 
$\Sigma_Q^*$-baryons having the spin-parities $J^P = 1/2^+$ 
and $J^P = 3/2^+$, respectively.
Inclusion of the $s$-quark increases the number of heavy
baryons in the multiplet which (in addition to the isospin)  
is characterized by the strangeness~$S$. 
If $S = -1$, there are two baryonic states~$\Xi_Q$ 
and~$\Xi_Q^\prime$ with $J^P = 1/2^+$ and $\Xi_Q^*$-baryon 
with $J^P = 3/2^+$. For $S = -2$, the baryons 
with $J^P = 1/2^+$ and $J^P = 3/2^+$ are called~$\Omega_Q$ 
and~$\Omega_Q^*$.

\begin{figure}[t]
\centering
\vspace{-0.1cm}
\includegraphics[width=10cm]{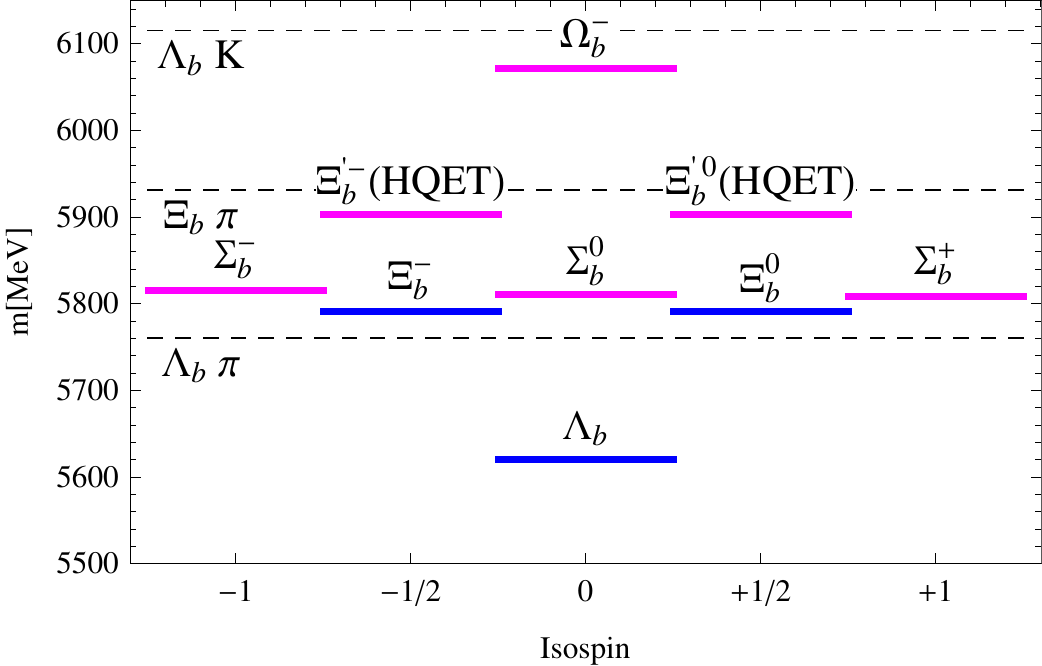}
\caption{
(Color online) Masses and thresholds of the $b$- baryons:
 $SU_F(3)$ triplet consists of $\Lambda_b$, $\Xi^-_b$, and $\Xi^0_b$ (in blue)
and the others belong to the $SU_F(3)$ sextet (in magenta).
The important hadronic thresholds $\Lambda_b \pi$, $\Xi_b \pi$ and 
$\Lambda_b K$ are indicated by horizontal dashed lines. The masses are listed in Tab.~\ref{tab:baryon-masses}. HQET calculations have been used in the case of lacking experimental input.
}
\vspace{-0.2cm}
\label{fig:SU(3)-multipletsMasses}
\end{figure}

The $SU(3)_{\rm F}$ multiplets of the ground-state bottom 
baryons are shown in Fig.~\ref{fig:SU(3)-multiplets}. 
We denote baryons from the triplet and sextet 
by $\baryongeneral{0}$ and $\baryongeneral{1}$, respectively. 
The corresponding masses of such baryons are presented in Tab.~\ref{tab:baryon-masses} 
and shown in Fig.~\ref{fig:SU(3)-multipletsMasses}. 
The important $\Lambda_b \pi$, $\Xi_b\pi$ and $\Lambda_b K$ thresholds 
are also specified in the figure. 
The spectrum of bottom baryons have been enlarged experimentally, 
substantially thanks to the effort done by the~CDF and~D0 
collaborations at the Tevatron collider during the last several years. 
In Tab.~\ref{tab:baryon-masses}, experimental measurements 
are taken from Ref.~\cite{Beringer:1900zz}, and we also compare 
with theoretical predictions (based on HQET~\cite{Liu:2007fg} 
and Lattice QCD~\cite{Lewis:2008fu}) for the masses of the 
ground-state bottom baryons (in units of MeV).  
Here, $\bar\Lambda = m_{H_b^j} - m_b$ 
and the continuum thresholds~$s_0$ in HQET for $m_b = 4.8$~GeV 
are also given (in units of GeV). The baryons lying below 
the corresponding strong decay thresholds $\Lambda_b \pi$, $\Xi_b \pi$ 
and $\Lambda_b K$ are of major importance, since the rare weak 
decays can be measured without the immense hadronic background.

The spinor of the baryon from the $j^p = 0^+$ triplet coincides 
with the spinor $u_\gamma$ of the heavy quark. 
In the case of the baryons from the $j^p = 1^+$ sextets 
the situation is more complicated. 
Both the Rarita-Schwinger vector-spinor $R^\mu_{\en\ga}(v)$ 
describing the $J^P = 3/2^+$ baryonic sextet
and the spinor of the baryon from the $J^P = 1/2^+$ sextet 
are specific products  
of the heavy-quark spinor $u_\ga(v)$ and the polarization 
vector~$\varepsilon^\mu(v)$ of the light diquark, in which $v^\mu$ is the four-velocity of the heavy baryon. 
This is the effect of spin decoupling of light and 
heavy degrees of freedom in the heavy quark limit.
The properties of~$u_\ga(v)$ and~$\varepsilon^\mu(v)$ are 
shortly reviewed below, while detailed information 
can be found in~\citeexplicit{Falk:1991nq}{Eqs. 2.12, 2.13}. 
Subsequently, the baryonic currents are constructed.

The spin and polarization sums are given by
\begin{align}\label{spinsumboth}
\sum\limits_{i=1}^2 u^i(v)\bar{u}^i(v)&=P_+\en\en\en
\tn{and}
\en\en\en
\sum\limits_{i=1}^3 \varepsilon^{*i}_{\mu}(v)\varepsilon^i_{\nu}(v)=-g_{\mu\nu}+v_{\mu}v_{\nu},
\end{align}
in which  $P_+=(1+\sla{v})/2$ and the normalization is given by
\begin{align}\label{hqnormalization}
-g^{\mu\nu}\sum\tn{Tr}\left[ u(v) \varepsilon^{*}_{\mu}(v)\varepsilon_{\nu}(v)   \bar{u}(v)      \right]&=6.
\end{align}
This corresponds to the number of degrees of freedom (3 from the light quark polarization vector,
since it has to fulfill the transversality condition $v_{\mu}\varepsilon^{\mu}(v)=0$
leaving 3 degrees of freedom, times 2 from the heavy quark  spinor). The polarization vector is normalized as $\varepsilon^{\mu}(v)\varepsilon_{\mu}^*(v)=1$.

\begin{table}[t]
\caption{Experimental measurements~\cite{Beringer:1900zz}
         and theoretical predictions 
         (based on HQET~\cite{Liu:2007fg} and Lattice 
         QCD~\cite{Lewis:2008fu}) for the masses of the 
         ground-state bottom baryons (in units of MeV).  
         Here, $\bar\Lambda = m_{H_b^j} - m_b$ and the 
         continuum thresholds~$s_0$ in HQET for  
         $m_b = 4.8$~GeV are also given (in units of GeV).}
\label{tab:baryon-masses} 
\begin{center} 
\begin{tabular}{ccccccc} 
\\ \hline  \hline
Baryon & $J^P$ & Experiment~\cite{Beringer:1900zz} & 
HQET~\cite{Liu:2007fg} & Lattice QCD~\cite{Lewis:2008fu} & 
$\bar\Lambda$ & $s_0$
\\ \hline  
$\Lambda_b$        & $1/2^+$ & $5619.4 \pm 0.7$ & 
                     $5637^{+68}_{-56}$ & 
                     $5641 \pm 21^{+15}_{-33}$ & 0.8 & 1.2 
\\[1mm] 
$\Sigma_b^+$       & $1/2^+$ & $5811.3 \pm 1.9$ & 
                     $5809^{+82}_{-76}$ & 
                     $5795 \pm 16^{+17}_{-26}$ & 1.0 & 1.3 
\\[1mm] 
$\Sigma_b^-$       & $1/2^+$ & $5815.5 \pm 1.8$ &  
                     $5809^{+82}_{-76}$ & 
                     $5795 \pm 16^{+17}_{-26}$ & 1.0 & 1.3 
\\[1mm] 
$\Sigma_b^{*+}$    & $3/2^+$ & $5832.1 \pm 1.9$ & 
                     $5835^{+82}_{-77}$ & 
                     $5842 \pm 26^{+20}_{-18}$ & 1.0 & 1.3 
\\[1mm] 
$\Sigma_b^{*-}$    & $3/2^+$ & $5835.1 \pm 1.9$ &  
                     $5835^{+82}_{-77}$ & 
                     $5842 \pm 26^{+20}_{-18}$ & 1.0 & 1.3 
\\[1mm]
$\Xi_b^-$          & $1/2^+$ & $5791.1 \pm 2.2$ & 
                     $5780^{+73}_{-68}$ & 
                     $5781 \pm 17^{+17}_{-16}$ & 1.0 & 1.3 
\\[1mm] 
$\Xi_b^0$          & $1/2^+$ & $5787.8 \pm 5.1$ & 
                     $5780^{+73}_{-68}$ & 
                     $5781 \pm 17^{+17}_{-16}$ & 1.0 & 1.3 
\\[1mm] 
$\Xi_b^\prime$     & $1/2^+$ &  & 
                     $5903^{+81}_{-79}$ & 
                     $5903 \pm 12^{+18}_{-19}$ & 1.1 & 1.4 
\\[1mm] 
$\Xi_b^{\prime *}$ & $3/2^+$ & & 
                     $5903^{+81}_{-79}$ & 
                     $5950 \pm 21^{+19}_{-21}$ & 1.1 & 1.4 
\\[1mm] 
$\Omega_b^-$       & $1/2^+$ & $6071 \pm 40$ & 
                     $6036  ^{+ 81}_{- 81}$      & 
                     $6006 \pm 10^{+20}_{-19}$ & 1.3 & 1.5 
\\[1mm] 
$\Omega_b^*$       & $3/2^+$ & & 
                     $6063^{+83}_{-82}$ & 
                     $6044 \pm 18^{+20}_{-21}$ & 1.3 & 1.5 
\\[1mm] \hline  \hline
\end{tabular} 
\end{center} 
\end{table}

The heavy quark spin has no fixed direction in the heavy quark limit and can be rotated at will. 
 The heavy baryon vector-spinor $\varepsilon^{\mu}(v)u_{\ga}(v)$
is therefore neither transforming as a spin $3/2$ nor as a spin $1/2$ quantity,
 since Lorentz invariance is explicitly broken.
One has to restore the proper Lorentz-invariant behavior by fixing the heavy quark
spinor with respect to the direction of the polarization vector $\varepsilon^{\mu}(v)$.
This can be done by using the transformation properties of a spin $3/2$
Rarita-Schwinger vector-spinor $\vectorspinor^{3/2\mu}_{\en\en\en\ga}(v)$ \citeexplicit{Falk:1991nq}{Eq. 2.9} and \citeexplicit{Grozin:2004yc}{p. 10}:
\index{Rarita-Schwinger vector-spinor}
\begin{align}\label{condfuerdreihalbe}
(\sla{v}\vectorspinor^{3/2\mu}(v))_{\ga}&=\vectorspinor^{3/2\mu}_{\en\en\en\ga}(v),\en\en\en v_{\mu}\vectorspinor^{3/2\mu}_{\en\en\en\ga}(v)=0, \en\en\en (\ga_{\mu}\vectorspinor^{3/2\mu}(v))_{\ga}=0.
\end{align}
Following  \citeexplicit{Falk:1991nq}{Eq. 2.10} we define
\begin{align}\label{eq_projectors}
(P_{3/2})_{\en\nu\ga}^{\mu\en\en\ga'}&=\left[\delta_{\nu}^{\mu}-\frac{1}{3}(\ga^{\mu}+v^{\mu})\ga_{\nu}\right]_{\ga}^{\en\ga'},\nonumber\\
(P_{1/2})_{\en\nu\ga}^{\mu\en\en\ga'}&=\left[\frac{1}{3}(\ga^{\mu}+v^{\mu})\ga_{\nu}\right]_{\ga}^{\en\ga'},
\end{align}
which fulfill the projection operator conditions $P^2=P$ and $P_{3/2}+P_{1/2}=\Id$. 
Now the product of the spinor $u_\gamma(v)$ and the polarization vector $\varepsilon^\mu(v)$ can be split in proper spin representations
\begin{align}
\varepsilon^{\mu}(v)u_{\ga}(v)=\vectorspinor^{3/2\mu}_{\en\en\en\ga}(v)+\vectorspinor^{1/2\mu}_{\en\en\en\ga}(v),
\end{align}
with
\begin{align}
\vectorspinor^{3/2\mu}_{\en\en\en\ga}(v)&= \left( \left[\varepsilon^{\mu}(v)-\frac{1}{3}(\ga^{\mu}+v^{\mu})\sla{\varepsilon}(v)\right] u(v) \right)_{\ga},  \nonumber\\
\vectorspinor^{1/2\mu}_{\en\en\en\ga}(v)&= \left( \left[\frac{1}{3}(\ga^{\mu}+v^{\mu})\sla{\varepsilon}(v)\right] u(v)\right)_{\ga}.
\end{align}
A further definition often used is
\begin{align}
\vectorspinor^{1/2\mu}(v)
\equiv
\frac{1}{\sqrt{3}}
(\gamma^\mu+v^\mu)
\gamma_5 \tilde R(v)
\end{align}
with
\begin{align}
\tilde \vectorspinor(v)
=
\frac{1}{\sqrt{3}}
\gamma_5\gamma_\mu
\vectorspinor^{1/2\mu}(v).
\end{align}
One can check that $\vectorspinor^{3/2\mu}_{\en\en\en\ga}(v)$ fulfills~\eqref{condfuerdreihalbe}, and $\vectorspinor^{1/2\mu}_{\en\en\en\ga}(v)$ is a spin $1/2$ spinor, with normalization
\begin{equation}
-g_{\mu\nu} \tn{Tr}\left[\bar{\vectorspinor}^{3/2\mu}(v)\vectorspinor^{3/2\nu}(v)\right]=4 \en\en \tn{and} \en\en -g_{\mu\nu} \tn{Tr}\left[\bar{\vectorspinor}^{1/2\mu}(v)\vectorspinor^{1/2\nu}(v)\right]
=
-\tn{Tr}\left[\bar{\tilde{\vectorspinor}}(v) \tilde\vectorspinor(v)\right]
=
2
,
\end{equation}
corresponding to~\eqref{hqnormalization}, where the $3/2$ vector-spinor gets 4 degrees of freedom and the $1/2$ spinor gets 2.
One should keep in mind, that this is a global rotation only to get the proper transformation properties. The rotation of the heavy quark spin will not affect the following calculations, which are performed in the heavy quark limit.

\section{Local interpolating operators of heavy baryons in HQET}
\label{sec_locop}

The most general three-quark current of the heavy baryon  
with one heavy quark in the heavy quark limit 
is given by the direct product 
of the diquark, constructed from two light quarks, and 
the heavy quark field~$Q$ due to spin decoupling of the 
heavy degrees of freedom, discussed in the last section.
In this limit, the decay constants $f^{(1)}_{\baryongeneral{0}}$ 
and $f^{(2)}_{\baryongeneral{0}}$ for the $j^p = 0^+$ triplet 
of bottom baryons are defined by the matrix elements 
of the local currents~\cite{Grozin:1992td,Groote:1997yr}:
\begin{eqnarray} 
\epsilon^{abc} \langle 0 | 
\left [ q_1^{a T} (0) {\cal C} \gamma_5 q_2^b (0) \right ] 
Q_\gamma^c (0) | \baryongeneral{0} (v) \rangle =  
f_{\baryongeneral{0}}^{(1)} u_\gamma (v) , 
\label{eq:f1-triplet-def} \\
\epsilon^{abc} \langle 0 | 
\left [ q_1^{a T} (0) {\cal C} \gamma_5 v\!\!\!\slash q_2^b (0) \right ] 
Q_\gamma^c (0) |\baryongeneral{0} (v) \rangle =  
f_{\baryongeneral{0}}^{(2)} u_\gamma(v) .  
\label{eq:f2-triplet-def}
\end{eqnarray} 
The decay constants $f^{(1)}_{\baryongeneral{1}}$ and 
$f^{(2)}_{\baryongeneral{1}}$ for the $j^p = 1^+$ 
bottom baryons are defined   
as 
\begin{eqnarray} 
\epsilon^{abc} \langle 0 | 
\left [ q_1^{a T} (0) {\cal C} \gamma_t^\nu q_2^b (0) \right ] 
Q_\gamma^c (0) | \baryongeneral{1} (v) \rangle =  
\frac{1}{\sqrt 3} \, f_{\baryongeneral{1}}^{(1)} 
\varepsilon^\nu (v) u_\gamma (v) ,  
\label{eq:f1-sextet-def} \\
\nonumber\\
\epsilon^{abc} \langle 0 | 
\left [ q_1^{a T} (0) {\cal C} \gamma_t^\nu 
v\!\!\!\slash q_2^b (0) \right ] 
Q_\gamma^c (0) | \baryongeneral{1} (v) \rangle = 
\frac{1}{\sqrt 3} \, f_{\baryongeneral{1}}^{(2)} 
\varepsilon^\nu (v) u_\gamma(v) ,
\label{eq:f2-sextet-def} 
\end{eqnarray} 
where $a, \ b, \ c = 1, 2, 3$ are the color indices, 
$\slashed v = v.\gamma$, 
$\gamma_t^\mu = \gamma^\mu - \slashed v \, v^\mu$, 
$Q$ is the effective static 
field of the heavy quark satisfying $\slashed v Q = Q$, 
the index~$T$ indicates a transposition, and
${\cal C}$ is the charge-conjugation matrix 
with the properties 
${\cal C} \gamma_\mu^T {\cal C}^{-1} = - \gamma_\mu$ 
and ${\cal C} \gamma_5 {\cal C}^{-1} = \gamma_5$.
In Eqs.~\eqref{eq:f1-sextet-def} and~\eqref{eq:f2-sextet-def},  
a factor $1/\sqrt 3$ is introduced so that the decay constants 
defined here agree with the ones  in~\cite{Grozin:1992td,Groote:1997yr}.
Note that the currents which are commonly used in the literature 
and define the decay constants, for example the two interpolating 
currents for the $\Sigma^*_b$-baryon~\cite{Groote:1996em}   
\begin{align} 
J^{\Sigma_b^*}_{1\mu} (x) & = \varepsilon^{a b c} \left [ 
q_1^{a T} (x) {\cal C} \gamma_t^\nu q_2^b (x) \right ] 
\left ( g_{\mu \nu} - \frac{1}{3} \, 
\gamma_{t \mu} \gamma_{t \nu} \right ) Q^c (x) ,
 \\  
J^{\Sigma_b^*}_{2\mu} (x) & = \varepsilon^{a b c} \left [ 
q_1^{a T} (x) {\cal C} \slashed v \gamma_t^\nu q_2^b (x) \right ] 
\left ( g_{\mu \nu}  - \frac{1}{3} \, 
\gamma_{t \mu} \gamma_{t \nu} \right ) Q^c (x) ,
\end{align}
can be easily derived from the currents defining the matrix 
elements in~\eqref{eq:f1-sextet-def} and~\eqref{eq:f2-sextet-def} 
by applying the projectors in~\eqref{eq_projectors}.

In QCD sum rule calculations of the LCDAs, the time-ordered product 
of the non-local and local currents (more precisely, 
the Dirac-conjugated local current) is involved.  
The most general structure of the Dirac-conjugated 
$j^p = 1^+$ local interpolating current 
(for the $j^p = 0^+$ current, see~\cite{Ball:2008fw}) 
is now given by the following linear combination:
\begin{equation}
\bar J_{\Gamma' \gamma} (x) = \varepsilon_{a b c} \left [ 
\bar q_1^a (x) \Gamma' {\cal C}^{-1} \bar q_2^{b T} (x) 
\right ]  
\bar Q_\gamma^c (x) ,
\label{thelocalcurrentdefinition}
\end{equation}
where $\Gamma' = \left ( A \Id + B \sla{v} \right ) \gamma_t^\mu$,
$B = 1 - A$, and $A \in [0,1]$. Following~\cite{Ball:2008fw}, 
the arbitrariness in the choice of the local current, {\it i.\,e.}, 
the variation in~$A$, will later be adopted as an error estimate, 
and we give the results for $A = 1/2$, which corresponds to the  
constituent quark model current that has maximal overlap with the 
ground-state baryons in the constituent quark model picture, 
in which all quarks are assumed to be on the mass 
shell~\citeexplicit{Groote:1997yr}{p. 5, Sec. 3.3}.

\section{Non-local light-cone operators of heavy baryons in HQET} 
\label{sec_nonlocop}

The currents used in the determination of the LCDAs 
are by definition non-local objects, aligned along 
a light-like direction~$n^\mu$, 
with coordinates $t_i n^\mu$ for the $i$-th quark.
For simplicity the baryon is defined in the frame 
of the $b$-quark $Q_\gamma = Q_\gamma (0)$, and 
the Wilson lines, which are needed to ensure gauge 
invariance of the current, are omitted. 
Their influence will be discussed whenever necessary.
Including the light-cone coordinate dependencies
one finds for the non-local light-cone operators 
that they can be written in the heavy quark limit as
\begin{equation}
\bra{\phantom{\baryongeneral{1}}\!\!\!\!\!\!\!\!\!\!\!\!\!0\;} 
\left [ q_1 (t_1) {\cal C} \Gamma^\mu q_2 (t_2) \right ] 
Q_\gamma (0) \ket{\baryongeneral{1} (v)}
= \frac{1}{\sqrt 3}f_{\baryongeneral{1}} \psi^\Gamma (t_1,t_2) \; 
\varepsilon^{ \mu} (v) u_\gamma (v) , 
\label{mostgeneralcurrentcoo}
\end{equation}
where the color indicies are suppressed to simplify the equation as well as the transposition superscript on $q_1(t_1)$. 
In~\eqref{mostgeneralcurrentcoo}, $\Gamma^\mu$ symbolizes all 
possible spinor structures allowed by the Lorentz symmetry and 
$\psi^{\Gamma} (t_1, t_2)$  are the light-cone distribution 
amplitudes corresponding to the current defined by~$\Gamma^\mu$. 
In total there are at maximum eight linear independent structures, 
$n^\mu \Id$, $\gamma^\mu$, $n^\mu \sla{v}$, $n^\mu \sla{n}$, 
$-i \sigma^{\mu \nu} v_\nu$, $-i\sigma^{\mu \nu} n_\nu$, 
$-i n^\mu \sigma^{\nu \rho} n_\nu v_\rho$ and 
$\epsilon^{\mu \nu \rho \tau} v_\nu n_\rho \gamma_\tau \gamma_5$ 
with $\epsilon^{0123} = 1$.

Since the quarks are aligned along the light-like direction $n^\mu$,
it is convenient for the description of the quark dynamics to work 
in the light-cone basis, where an arbitrary vector~$a^\mu$ 
is decomposed as 
\index{Light-cone coordinates}
\begin{align}
a^\mu &= \frac{1}{2} 
\left ( a_- n^\mu + a_+ \bar{n}^\mu \right ) + a_\perp^\mu,  
\label{eq:vecror-decomp}  
\end{align}
in which $n^\mu = (1,0,0,-1)$ and $\bar{n}^\mu = (1,0,0,1)$ 
are light-cone vectors with $n^2 = \bar{n}^2 = 0$, 
$\produkt{n}{\bar{n}} = 2$, and $a_\perp^\mu$ refers to the 
remaining two space-like dimensions, which are perpendicular 
to both~$n^\mu$ and~$\bar{n}^\mu$. The scalar product of two 
vectors~$a^{\mu}$ and~$b^{\mu}$ is given by
\begin{align}
\produkt{a}{b} & = \frac{1}{2} 
\left ( a_+ b_- + a_- b_+ \right ) + a_{\perp \mu} b_\perp^\mu.  
\label{eq:vector-product}
\end{align}
The coefficients of the decomposed kinematical vectors
\begin{align}
v^\mu           & = \frac{1}{2} 
\left ( \frac{n^\mu}{v_+} + v_+ \bar{n}^\mu \right ) , 
\label{eq:v-decomp} \\
\varepsilon^\mu (v) & = \frac{e}{2} 
\left ( \frac{n^\mu}{v_+} - v_+ \bar{n}^\mu \right ) + 
\varepsilon_\perp^\mu (v) ,
\label{eq:eps-decomp}
\end{align}
have been chosen in a way to fulfill the conditions $v^2 = 1$, 
$\varepsilon^\mu (v) \varepsilon_\mu (v) = -1$ and 
the transversality condition $v_\mu \varepsilon^\mu (v) = 0$.
The light-like vertors are chosen in a way that~$v^\mu$ 
has no perpendicular components.  Accordingly, 
$\varepsilon_\perp^\mu (v)$ and $\varepsilon_\parallel^\mu (v) 
\equiv \varepsilon^\mu (v) - \varepsilon_\perp^\mu (v)$ 
are called transverse and parallel polarizations, respectively.
They fulfill the conditions 
$\varepsilon_{\parallel\mu} (v) \varepsilon_\perp^\mu (v) = 0$, 
$\varepsilon_{\parallel\mu} (v) \varepsilon_\parallel^\mu (v) = - e^2$,
$\varepsilon_{\perp\mu} (v) \varepsilon_\perp^\mu (v) = e^2 - 1$ 
and $n_\mu \varepsilon^\mu (v) = - e v_+$ (in the following, 
the reference of the polarization vector and its components 
to the four-velocity~$v$ will be understood implicitly).  
The scalar variable~$e$ is the measure for the amount 
of the diquark parallel polarization. Note that the index~$\perp$ 
refers to being perpendicular to both vectors~$n^\mu$ and~$\bar n^\mu$ 
(and also the four-velocity~$v^\mu$), while the index~$t$ refers 
to being perpendicular to~$v^\mu$ only.
The vector 
\begin{equation} 
\bar v^\mu = \frac{1}{2} \left ( 
\frac{n^\mu}{v_+} - v_+ \bar{n}^\mu 
\right ) 
\label{eq:vbar-def}
\end{equation} 
is the only possible normalized combination ($\bar v^2 = -1$) 
of~$n^\mu$ and $\bar n^\mu$ which fulfills the transversality 
condition $v_\mu \bar v^\mu = 0$. In this notation, 
the parallel polarization vector is simply given by
$\varepsilon_\parallel^\mu  = e \, \bar v^\mu$.

For comparison, let us recall the set of the non-local currents 
for the spin-parity $j^p = 0^+$ diquark and the corresponding 
LCDAs given in~\cite{Ball:2008fw} with $\{ \psi^n, \,
\psi^{n \bar n}, \, \psi^\Id, \, \psi^{\bar n} \}$
corresponding to $\{ \Psi_2, \, \Psi_3^\sigma, \,
\Psi_3^s, \, \Psi_4 \}$ and a change of the labeling of the decay constants, which are labeled here by the superscript $(1)$ for vector and $(2)$ for tensor currents:
\begin{align} 
\frac{1}{v_+} 
\bra{\phantom{\baryongeneral{1}}\!\!\!\!\!\!\!\!\!\!\!\!\!0\;} 
\left [ q_1 (t_1) {\cal C} \gamma_5 \sla{n} q_2 (t_2) \right ] 
Q_\gamma \ket{\baryongeneral{0}} 
& =
\psi^{n} (t_1, t_2) \, f^{(1)}_{\baryongeneral{0}} u_\gamma ,
\nonumber \\
\frac{i}{2} 
\bra{\phantom{\baryongeneral{1}}\!\!\!\!\!\!\!\!\!\!\!\!\!0\;} 
\left [ q_1 (t_1) {\cal C} \gamma_5 \sigma_{\bar{n} n} q_2 (t_2) \right ]  
Q_\gamma \ket{\baryongeneral{0}} 
& =
\psi^{n\bar{n}} (t_1, t_2) \, f^{(2)}_{\baryongeneral{0}} u_\gamma ,
\nonumber\\
\bra{\phantom{\baryongeneral{1}}\!\!\!\!\!\!\!\!\!\!\!\!\!0\;} 
\left [ q_1 (t_1) {\cal C} \gamma_5 q_2 (t_2) \right ]  
Q_\gamma \ket{\baryongeneral{0}} 
& = 
\psi^{\Id} (t_1, t_2) \, f^{(2)}_{\baryongeneral{0}} u_\gamma , 
\nonumber \\
v_+ \bra{\phantom{\baryongeneral{1}}\!\!\!\!\!\!\!\!\!\!\!\!\!0\;} 
\left [ q_1 (t_1) {\cal C} \gamma_5 \sla{\bar{n}} q_2 (t_2) \right ]  
Q_\gamma \ket{\baryongeneral{0}} 
& =  
\psi^{\bar{n}} (t_1, t_2) \, f^{(1)}_{\baryongeneral{0}} u_\gamma , 
\label{spin0currents}
\end{align} 
where $\sigma_{\bar{n} n} = \sigma_{\mu \nu} \bar n^\mu n^\nu$.    
In a similar way, we introduce the set of the parallel 
currents and their LCDAs as follows: 
\index{Interpolating operators non-local $b$-baryon}
\begin{align} 
\frac{\bar v^\mu}{v_+} 
\bra{\phantom{\baryongeneral{1}}\!\!\!\!\!\!\!\!\!\!\!\!\!0\;} 
\left [ q_1 (t_1) {\cal C} \sla{n} q_2 (t_2) \right ]  
Q_\gamma \ket{\baryongeneral{1}} 
& =  \frac{1}{\sqrt 3}
\psi_\|^{n} (t_1,t_2) \, f^{(1)}_{\baryongeneral{1}} \, 
\varepsilon_\parallel^\mu u_\gamma ,
\nonumber \\
\frac{i \bar v^\mu}{2} 
\bra{\phantom{\baryongeneral{1}}\!\!\!\!\!\!\!\!\!\!\!\!\!0\;} 
\left [ q_1 (t_1) {\cal C} \sigma_{\bar{n} n} q_2 (t_2) \right ]  
Q_\gamma \ket{\baryongeneral{1}} 
& = \frac{1}{\sqrt 3}
\psi_\|^{n\bar{n}} (t_1,t_2) \, f^{(2)}_{\baryongeneral{1}} \, 
\varepsilon_\parallel^\mu u_\gamma ,
\nonumber \\
\bar v^\mu 
\bra{\phantom{\baryongeneral{1}}\!\!\!\!\!\!\!\!\!\!\!\!\!0\;} 
\left [ q_1 (t_1) {\cal C} q_2 (t_2) \right ]  
Q_\gamma \ket{\baryongeneral{1}} 
& =  \frac{1}{\sqrt 3}
\psi_\|^{\Id} (t_1,t_2) \, f^{(2)}_{\baryongeneral{1}} \, 
\varepsilon_\parallel^\mu u_\gamma, 
\nonumber \\
-v_+ \bar v^\mu  
\bra{\phantom{\baryongeneral{1}}\!\!\!\!\!\!\!\!\!\!\!\!\!0\;}
\left [ q_1 (t_1) {\cal C} \sla{\bar{n}} q_2 (t_2) \right ]  
Q_\gamma \ket{\baryongeneral{1}} 
& =  \frac{1}{\sqrt 3}
\psi_\|^{\bar{n}} (t_1,t_2) \, f^{(1)}_{\baryongeneral{1}} \, 
\varepsilon_\parallel^\mu u_\gamma ,
\label{parallelcurrents}
\end{align}
and the set of the transverse currents and LCDAs: 
\begin{align}
\frac{1}{v_+} 
\bra{\phantom{\baryongeneral{1}}\!\!\!\!\!\!\!\!\!\!\!\!\!0\;}
\left [ q_1 (t_1) {\cal C} \slashed{n} \gamma_\perp^\mu q_2 (t_2) \right ]  
Q_\gamma \ket{\baryongeneral{1}} 
& = \frac{1}{\sqrt 3}
\psi_\perp^n (t_1,t_2) \, f^{(2)}_{\baryongeneral{1}} \,  
\varepsilon_\perp^\mu u_\gamma , 
\nonumber \\
\bra{\phantom{\baryongeneral{1}}\!\!\!\!\!\!\!\!\!\!\!\!\!0\;} 
\left [ q_1 (t_1) {\cal C} \gamma_\perp^\mu q_2 (t_2) \right ]  
Q_\gamma \ket{\baryongeneral{1}} 
& =  \frac{1}{\sqrt 3}
\psi_\perp^\Id (t_1,t_2) \, f^{(1)}_{\baryongeneral{1}} \, 
\varepsilon_\perp^\mu u_\gamma , 
\nonumber \\
\frac{i}{2}
\bra{\phantom{\baryongeneral{1}}\!\!\!\!\!\!\!\!\!\!\!\!\!0\;}
\left [ q_1 (t_1) {\cal C} \sigma_{\bar{n} n} \gamma_\perp^\mu q_2 (t_2) \right ]  
Q_\gamma \ket{\baryongeneral{1}} 
& = \frac{1}{\sqrt 3}
\psi_\perp^{n\bar n} (t_1,t_2) \, f^{(1)}_{\baryongeneral{1}} \, 
\varepsilon_\perp^\mu u_\gamma, 
\nonumber \\
v_+ 
\bra{\phantom{\baryongeneral{1}}\!\!\!\!\!\!\!\!\!\!\!\!\!0\;}
\left [ q_1 (t_1) {\cal C} \bar{\slashed{n}} \gamma_\perp^\mu q_2 (t_2) \right ]   
Q_\gamma \ket{\baryongeneral{1}} 
& = \frac{1}{\sqrt 3}
\psi_\perp^{\bar{n}} (t_1,t_2) \, f^{(2)}_{\baryongeneral{1}} \, 
\varepsilon_\perp^\mu u_\gamma . 
\label{transversalcurrents}
\end{align}
The notations $\gamma_\perp^\mu =
\gamma^\mu - \frac{1}{2} \left ( 
\sla{n} \bar{n}^\mu + \sla{\bar{n}} n^\mu \right )$,
$\{ \gamma_\perp^\mu, \sla{n} \} = 
 \{ \gamma_\perp^\mu, \sla{\bar n} \} = 0$ have been used. 
All currents inherit the transversality property 
from the polarization vector.  
Note, that there is no interpolating current 
involving $\gamma_\parallel^\mu$, 
since $\gamma_\parallel^\mu = - \sla{\bar v} \bar v^\mu$.  
As a result, such currents are linear combinations of the parallel 
currents defined in~\eqref{parallelcurrents}.
In a more compact form, the operators can be written as follows: 
\begin{align}
&\hspace{-0.5cm}\bra{\phantom{\baryongeneral{1}}\!\!\!\!\!\!\!\!\!\!\!\!\!0\;} q_1(t_1 n)
{\cal C}
\Gamma
q_2(t_2 n)Q_{\ga}
 \ket{\baryongeneral{1}}
\nonumber\\
&=
\frac{1}{4 \sqrt 3}{\rm Tr}\bigg[
\Gamma
.
\bigg(
f^{(2)}_{\baryongeneral{1}} \frac{i}{2} 
\left( v_{+}  \sigma_{\bar n \varepsilon_{\perp}} \psi^{ n}_\perp (t_1,t_2) 
+e \sigma_{n\bar n } \psi^{n\bar{n}}_\parallel (t_1,t_2) 
+\frac{1}{v_{+}}\sigma_{n \varepsilon_{\perp}  }\psi^{\bar{n}}_\perp  (t_1,t_2)
\right)
\nonumber\\
&\quad
+ 
e  
f^{(2)}_{\baryongeneral{1}} \psi^{\Id}_\parallel (t_1,t_2) 
+
\frac{i}{2} 
f^{(1)}_{\baryongeneral{1}}  \gamma_5\gamma_\alpha
\epsilon^{\alpha\varepsilon_\perp n \bar{n}}
 \psi^{n\bar{n}}_\perp (t_1,t_2)
\nonumber\\
&\quad
+
f^{(1)}_{\baryongeneral{1}} 
\left(
\frac{e\; v_+}{2}
\slashed{\bar n}\psi^{n}_\parallel (t_1,t_2)
+ \slashed{\varepsilon}_{\perp} \psi^{\Id}_\perp (t_1,t_2)
- \frac{e}{2 v_+}\slashed{n} \psi^{\bar{n}}_\parallel (t_1,t_2)
\right)
\bigg)
\bigg ] u_\gamma ,
\label{covtrace}
\end{align}
in which $\Gamma$ represents a Dirac matrix and the decay constants 
$f^{(1)}_{\baryongeneral{1}}$ and $f^{(2)}_{\baryongeneral{1}}$  
correspond to the vector and tensor currents, respectively.  
Using~\eqref{covtrace}, it is straightforward to calculate 
the relations of the wave functions in any given operator basis.
The operators in~\eqref{eq_projectors} for the restoration 
of Lorentz invariance can be applied to the currents 
in~\eqref{spin0currents}~---~\eqref{transversalcurrents} 
to yield the proper interpolating currents used in, {\it e.\,g.}, 
light-cone sum rule calculations of form factors.

The twist is given by an expansion in~$v_+$  
(see for comparison~\citeexplicit{Grozin:1996pq}{Eq. 2.5}).
The leading, subleading and subsubleading twists are 
proportional to~$1/v_+$,~1 and~$v_+$, respectively.
Similar to the case of light vector mesons 
(see, for example,~\citeexplicit{Ball:1998sk}{Tab. 2}) 
one obtains the twist in the baryon case, listed in Tab.~\ref{tab:BaryonTwist}. 
\begin{table}[h]
\caption{Twist of the baryon LCDAs, defined in Eqs.~\eqref{spin0currents} to~\eqref{transversalcurrents}.}
\label{tab:BaryonTwist} 
\begin{center} 
\def\arraystretch{1.5}%
\begin{tabular}{c|c|c|c}
 \hline \hline
\tn{twist:}       & 2             & 3                                  & 4                   \\\hline
\tn{scalar}    & $\psi^n$     & $\psi^\Id,   \psi^{n\bar{n}}$  & $\psi^{\bar{n}}$   \\
\tn{parallel}    &$ \psi_\|^n   $  &$ \psi_\|^\Id,   \psi_\|^{n\bar{n}} $ &$ \psi_\|^{\bar{n}}$   \\
\tn{transverse} &$ \psi_\perp^n $ & $\psi_\perp^\Id, \psi_\perp^{n\bar{n}} $      & $\psi_\perp^{\bar{n}}$ \\
 \hline \hline
\end{tabular} 
\end{center} 
\end{table}

The argument for the twist ordering is the following. 
Based on the choice of basis for~$v^\mu$, 
the orientation of the currents with respect to~$v^\mu$ induces 
a kinematical difference. The light-cone vectors~$\bar{n}^\mu$ 
and~$n^\mu$ differ only in their direction in the spacial 
coordinates along the $z$-direction. 
Hence, the characteristic feature is, with which three-velocity 
the baryon moves along the $z$-axis in the spacial direction of $n$ or $\bar n$.
In conclusion, the importance (or twist) can be characterized by~$v_+$.
However, this is not necessarily the conventional definition 
of ``twist$=$dimension$-$spin''.

The symmetric LCDAs  $\psi_{\|,\perp}^n (t_1, t_2)$, 
$\psi_{\|,\perp}^\Id (t_1, t_2)$ and
$\psi_{\|,\perp}^{\bar n} (t_1, t_2)$ defined in~\eqref{parallelcurrents} 
and~\eqref{transversalcurrents} (similar for \eqref{spin0currents}) are normalized to give 
the corresponding decay constants~$f^{(1)}_{\baryongeneral{1}}$ 
and~$f^{(2)}_{\baryongeneral{1}}$ in the local limit,
in which they are satisfying $\psi_{\|, \perp}^{n, \Id, \bar n} (0, 0) \equiv 1$.  
Performing a Fourier transform ($t_i \rightarrow \omega_i$ and 
$\omega_1 = u \omega$, $\omega_2 = \bar u \omega$ with $\bar u = 1 - u$) 
leads to the following normalizations:
\begin{equation}
\int\limits_0^\infty d\omega_1 \int\limits_0^\infty d\omega_2 \, 
\psi^{n, \Id, \bar n}_{\parallel,\perp} (\omega_1,\omega_2) =
\int\limits_0^\infty \omega d\omega \int\limits_0^1 du \, 
\psi^{n, \Id, \bar n}_{\parallel,\perp} (\omega,u) = 1.
\label{normalizationinlocallimit}
\end{equation}
The antisymmetric LCDAs $\psi_{\|, \perp}^{n \bar n} (t_1, t_2)$ are correspondingly normalized by their first Gegenbauer  moments:
\begin{equation}
\int\limits_0^\infty \omega d\omega 
\int\limits_0^1 du \, 
(2u -1)
\psi^{n\bar n}_{\parallel, \perp} (\omega,u) = 1,
\label{normalizationinlocallimit2}
\end{equation}
see Sec.~\ref{lcdamodel} for details of the model functions.
Here, $\omega_i$ are the light-like momenta of the light quarks~$q_i$, 
$\omega = \omega_1 + \omega_2$ is the total light-like momentum of
the light diquark system and~$u$ and~$\bar u$ are the momentum 
fractions of the light quarks.
Eqs.~\eqref{normalizationinlocallimit} and~\eqref{normalizationinlocallimit2}  indicate that the LCDAs are 
constructed in a way, such that the energy scale is given by the 
constants $f^{(1)}_{\baryongeneral{1}}$ and $f^{(2)}_{\baryongeneral{1}}$, 
while the coordinate dependencies are kept in the functions~$\psi$. 
Hence, the two decay constants are determined by the interpolating 
operators~\eqref{parallelcurrents} and~\eqref{transversalcurrents} 
in the local limit, {\it i.\,e.}, by the ones given 
in Eqs.~\eqref{eq:f1-sextet-def} and~\eqref{eq:f2-sextet-def}. 
Splitting the energy dependence from the coordinate dependencies 
leads to the advantage that the decay constants can be calculated 
at higher orders than the more complicated distribution amplitudes.

\section{Renormalization and Scale-Dependence of the Matrix Elements} 
\label{sec_renorm}

A non-relativistic constituent quark picture 
of the $\baryongeneralW$-baryon suggests that 
$f^{(2)}_{\baryongeneralW} \simeq f^{(1)}_{\baryongeneralW}$ 
at low scales of order 1~GeV, and this expectation is supported 
by numerous QCD sum rule calculations~\cite{%
Grozin:1992td,Groote:1997yr,Groote:1996em,Bagan:1993ii}.
In fact, the difference between the two couplings is only 
obtained at the level of NLO perturbative corrections 
to the sum rules~\cite{Groote:1997yr,Groote:1996em} 
and is numerically small.

The scale dependence of the couplings is given by 
\begin{equation}
f^{(i)}_{\baryongeneralW} (\mu) = 
f^{(i)}_{\baryongeneralW} (\mu_0) \left( 
\frac{\alpha_s(\mu)}{\alpha_s(\mu_0)} 
\right)^{\gamma^{(i)}_1/\beta_0} \left [ 
1 - \frac{\alpha_s(\mu_0) - \alpha_s(\mu)}{4\pi} \, 
\frac{\gamma^{(i)}_1}{\beta_0} \left ( 
\frac{\gamma^{(i)}_2}{\gamma^{(i)}_1} - \frac{\beta_1}{\beta_0} 
\right ) \right], 
\label{eq:f-scale-depen}
\end{equation} 
where the anomalous dimensions of the local interpolating operators  
\begin{equation}
\frac{d\ln f_{\baryongeneralW}^{(i)}(\mu)}{d\ln \mu} \equiv 
-\gamma^{(i)} = - \sum_k \gamma^{(i)}_k \, a^k (\mu) , 
\qquad 
a(\mu) \equiv \frac{\alpha_s^{\overline{\rm MS}}(\mu)}{4\pi},   
\label{eq:ADf-def}
\end{equation}
are known to NLO~\cite{Groote:1996em}, 
and the $\beta$-function expansion 
\begin{equation} 
\frac{d a(\mu)}{d\ln \mu} = 
- \beta_0 \, a^2 (\mu) - \beta_1 \, a^3 (\mu) + \cdots , 
\label{eq:Beta-func-def}
\end{equation}
was used to NLO with the coefficients 
$\beta_0 = 2 \left ( 11 - 2 n_f/3 \right )$ and 
$\beta_1 = 4 \left ( 51 - 19 n_f/3 \right )$.

The decay constants $f_{\Lambda_b} \approx 0.03 \; \tn{GeV}^3$ and 
$f_{\Sigma_b}\approx 0.038\; \tn{GeV}^3$ are known at NLO in the 
$SU(3)_{\rm F}$ limit~\citeexplicit{Groote:1997yr}{p. 1, Tab. 2}, 
while the $SU(3)_{\rm F}$ breaking effects are only known at~LO.
The decay constants $f^{(1)}_{\baryongeneral{1}}$ and 
$f^{(2)}_{\baryongeneral{1}}$ coincide approximately 
at the renormalization scale $\mu \approx 1$~GeV~\cite{Ball:2008fw}.
Note that these couplings cannot coincide 
at all scales since the corresponding operators 
have different anomalous dimensions.

\begin{figure}[t]
\centering
\includegraphics[width=0.99\textwidth]{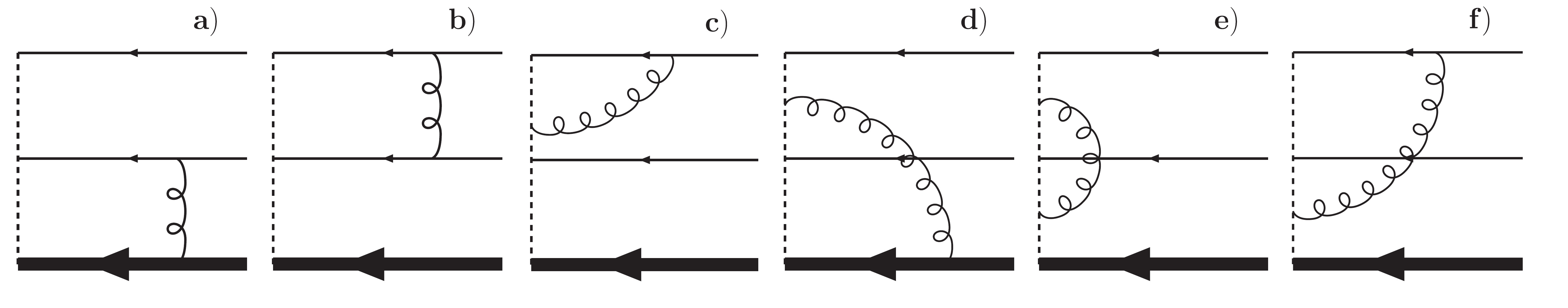}
\caption{
Types of diagrams for the non-local vertex renormalization. 
Each of the diagrams a), c), d) and f) appears twice (one for 
each light quark). The thick and thin solid lines correspond  
to the heavy and light quarks, respectively, and  
the dashed line represents the gauge link (Wilson line).
\label{vertexrenormalization}}
\end{figure}

The LCDAs are defined via the relations~\eqref{spin0currents}
to~\eqref{transversalcurrents}.
Thus they are not directly measurable quantities and 
obey renormalization group equations. 
They can be viewed as non-local three-quark vertices, 
just as the local interpolating currents can be viewed 
as local vertices. The gauge links, also called the 
Wilson lines, take account for the non-locality. 
Up to a first order expansion of the Wilson lines in~$g_s$, 
the non-local vertices are given by:
\begin{eqnarray}
\raisebox{-21pt}{
\includegraphics[width=0.25\textwidth]{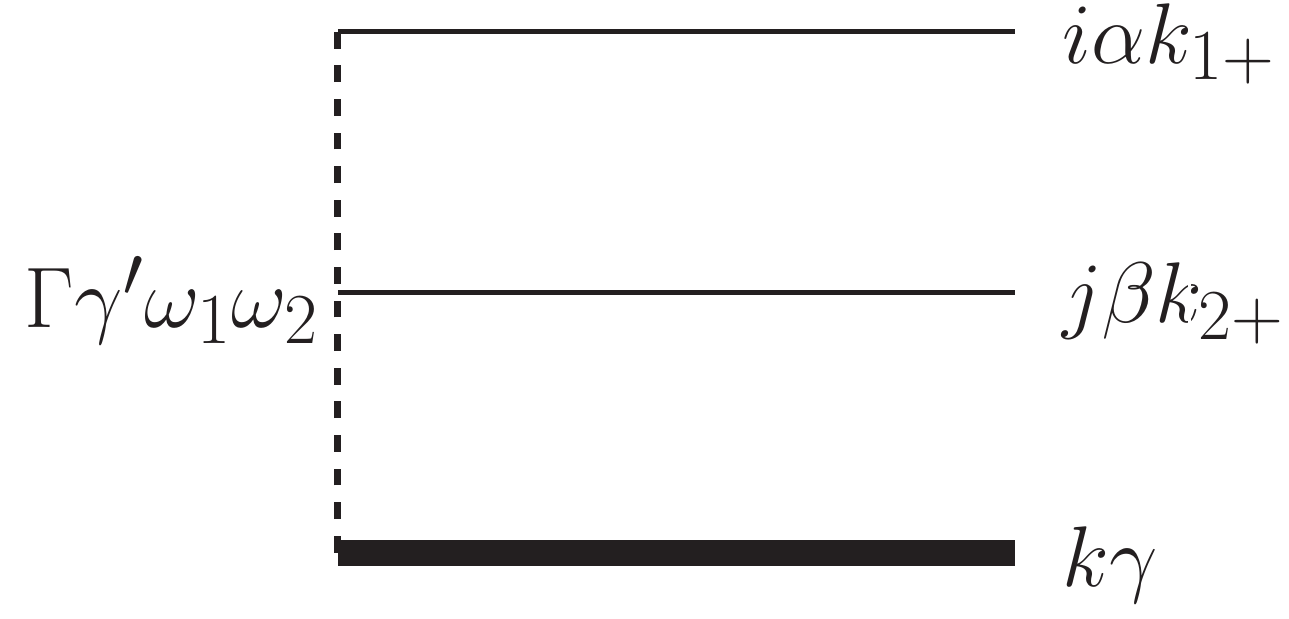}} 
& = & 
\begin{array}{l} 
-\epsilon_{ijk} \Gamma^{\alpha\beta} \Id_\gamma^{\en\gamma'} 
\delta ( \omega_1 - k_{1+} ) \delta ( \omega_2 - k_{2+} ), 
\end{array}
\nonumber \\
\nonumber \\
\raisebox{-25pt}{
\includegraphics[width=0.25\textwidth]{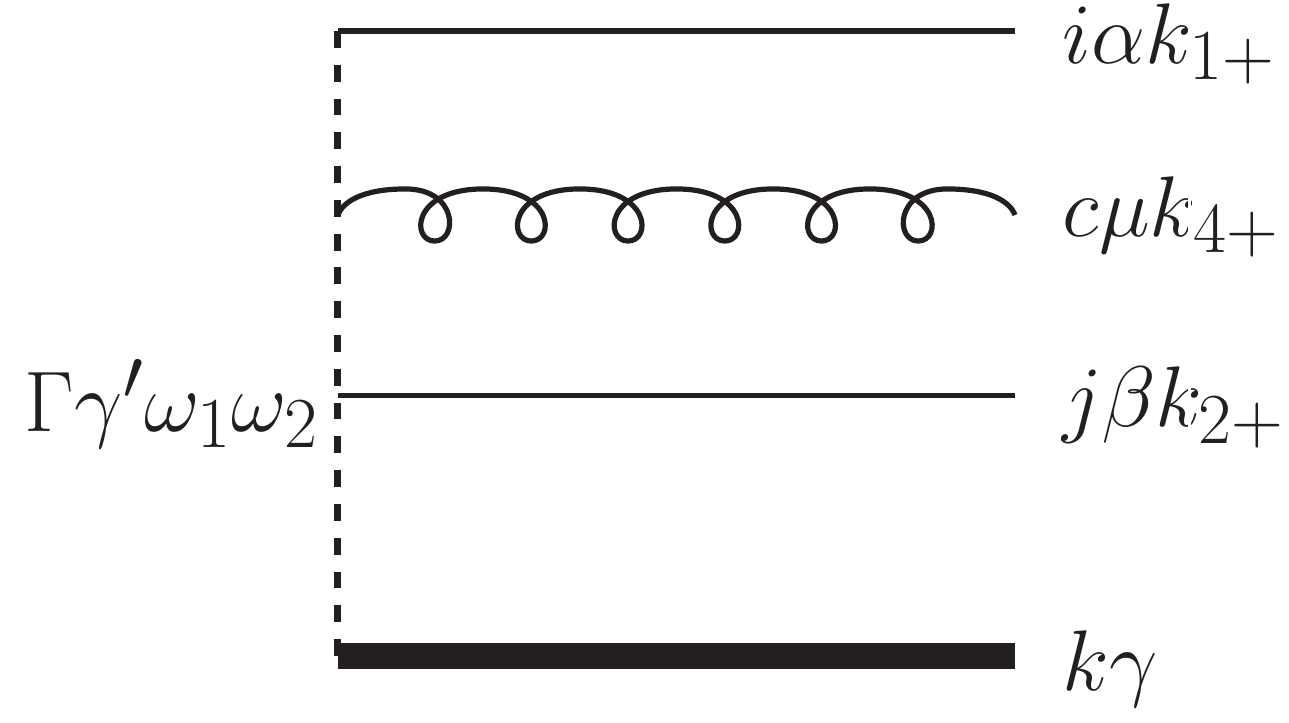}} 
& = & 
\begin{array}{l}
-g_s \epsilon_{ljk} t^{c l}_{\en\en i} n^\mu \Gamma^{\alpha\beta} 
\Id_\gamma^{\en\gamma'} \frac{1}{k_{4+}} 
\\ \\ 
\delta ( \omega_2 - k_{2+} ) \left [ 
\delta( \omega_1 - k_{1+} ) - \delta ( \omega_1 - k_{4+} - k_{1+} )        
\right ],
\end{array}
\label{eq_vertrendef}
\end{eqnarray}
in which color ($i,j,\ldots$) and spinor ($\alpha,\beta,\ldots$) indices are explicitly shown together with the light cone momenta $\omega_i, k_{i+}$ and the heavy quark momentum $k$.
The vertex for the case, in which the gluon line is attached 
to the lower Wilson line is obtained by replacing $1 \to 2$. 
The expansion of the Green's functions in~$g_s$ corresponds 
to an expansion of the Wilson lines in~$g_s$. In the $n$-th 
order expansion term, there are in general~$n$ gluon fields 
attached to the vertex, but one-loop calculations are sufficient for the computations in this work, and additional vertices 
to the ones in Eq.~\eqref{eq_vertrendef} are not necessary. 
The evolution kernel at one-loop order is determined by the 
ultraviolet poles of the matrix elements of the bare operators.
The types of diagrams which contribute to the kernel of the 
renormalization group equation up to one-loop order are given 
in Fig.~\ref{vertexrenormalization}.
For the leading twist (neglecting masses) the diagrams give 
the same contribution for the $SU(3)_{\rm F}$ triplet and 
sextet (Fig.~\ref{fig:SU(3)-multiplets}), where only for 
the transverse distribution amplitude diagram~{\bf b)} vanishes. 
The $SU(3)_{\rm F}$ triplet evolution equation has already 
been derived in~\citeexplicit{Ball:2008fw}{Eq. 13}.  

At one-loop order the diagrams shown in Fig.~\ref{vertexrenormalization} 
involve at the maximum two quarks in the loop. Hence, the derivation of the evolution kernels is almost 
identical to the meson case. The heavy-light contributions, involving 
the $b$-quark and one light quark, are similar to the heavy-light 
mesons~\cite{Lange:2003ff} and the light-light contributions are similar 
to the light-light mesons~\cite{Ball:2003sc}.
The diagram~{\bf a)} is UV finite as it is found in~\cite{Lange:2003ff}. 
The diagram~{\bf e)} vanishes since the corresponding vertex is 
proportional to $n^\mu n^\nu$ and the gluon propagator is proportional 
to $g_{\mu\nu}$. The evolution equation for the leading twist 
reads~\cite{Ball:2008fw}: 
\begin{eqnarray}
& &
\mu \, \frac{d}{d\mu} \, \psi^n (\omega_1, \omega_2; \mu) = 
- \frac{\alpha_s (\mu)}{2 \pi} 
\left ( 1 + \frac{1}{N_c} \right ) \left [
\int\limits_0^\infty d k_{1+} \, 
\gamma^{\rm LN} (\omega_1, k_{1+}; \mu) \, 
\psi^n (k_{1+}, \omega_2; \mu)
\right.
\nonumber \\
&&
\hspace*{10mm} 
\left. +
\int\limits_0^\infty d k_{2+} \, 
\gamma^{\rm LN} (\omega_2, k_{2+}; \mu) \, 
\psi^n (\omega_1, k_{2+}; \mu)
- \int\limits_0^1 d v \, V (u, v) \, 
\psi^n (v \omega, \bar v \omega; \mu)
+ \frac{3}{2} \, \psi^n (\omega_1, \omega_2; \mu)
\right ] ,
\label{eq_fullevolutionequation}
\end{eqnarray}
in which
\begin{eqnarray}
\gamma^{\rm LN} (\omega, k_+; \mu)
&=&
\ln \frac{\mu}{\omega} - \frac{5}{4}
-
\left [
\frac{\omega}{k_+} \, 
\frac{\Theta(k_+ - \omega)}{k_+ - \omega} +
\frac{\Theta(\omega - k_+)}{\omega - k_+}
\right ]_\oplus 
\label{eq:LN-kernel}
\end{eqnarray}
is taken from~\citeexplicit{Lange:2003ff}{Eq. 8} and $V (u,v)$ is the well-known
Efremov-Radyushkin-Brodsky-Lepage kernel~\cite{Efremov:1978rn,Lepage:1979zb}:
\begin{eqnarray}
V (u,v) &=&
\left [
\frac{1 - u}{1 - v} 
\left( 1 + \frac{1}{u - v} \right ) \Theta ( u - v )
+
\frac{u}{v}
\left( 1 + \frac{1}{v - u} \right ) \Theta ( v - u )
\right ]_+,  
\label{eq:ERBL-kernel}
\end{eqnarray} 
For transverse leading-twist LCDA $\psi_\perp^n$ diagram~{\bf b)} 
vanishes and one needs to replace
\begin{eqnarray}
V (u,v) & \to &
V(u,v) -
\left [ 
\frac{1 - u}{1 - v} \, \Theta(u - v) +
\frac{u}{v} \, \Theta(v - u)
\right]_+,  
\label{eq:ERBL-kernel-trans}  
\end{eqnarray}
while the rest stays unchanged. 
This effect on the evolution is negligible.
The~``$\oplus$'' and~``$+$'' subtractions 
are defined as follows: 
\begin{eqnarray}
&& 
\int\limits_0^\infty d k_+
\left [ \gamma (\omega, k_+) \right]_\oplus f (k_+) =
\int\limits_0^\infty d k_+
\gamma (\omega, k_+) \left [ f (k_+) - f (\omega) \right ], 
\label{eq:oplus-def} \\
&& 
\left [ V (u, v) \right ]_+ =
V (u, v) - \delta (u - v) \int\limits_0^1 dt \, V (t,v) . 
\label{eq:plus-def}
\end{eqnarray}
For small evolution steps, $\ln (\mu/\mu_0) \lesssim 1$, 
the differentiation with respect to~$\mu$ in 
Eq.~\eqref{eq_fullevolutionequation} is given 
by the linear approximation: 
\begin{equation}
\mu \, \frac{d}{d\mu} \, \psi^n (\omega_1,\omega_2; \mu)
\approx
\frac{\psi^n (\omega_1, \omega_2; \mu) - \psi^n (\omega_1, \omega_2; \mu_0)}
     {\ln (\mu/\mu_0)},
\label{eq:RGE-solve-approx}
\end{equation}
and the evolution of the distribution amplitudes 
can be calculated easily by substituting the initial condition 
$\psi^n (\omega_1, \omega_2; \mu_0)$ in the integrand 
in~\eqref{eq_fullevolutionequation}.
In the following we give the parallel LCDAs as example. 
As will be shown, the effect of the renormalization is within 
the errors obtained from the variation of~$A$ characterizing 
two different structures in the local interpolating currents.

\section{QCD Sum Rules}
\label{sec_corrfun} 

\begin{figure}
\centering
\resizebox{0.54\textwidth}{0.2\textwidth}{
\includegraphics[width=0.55\textwidth]{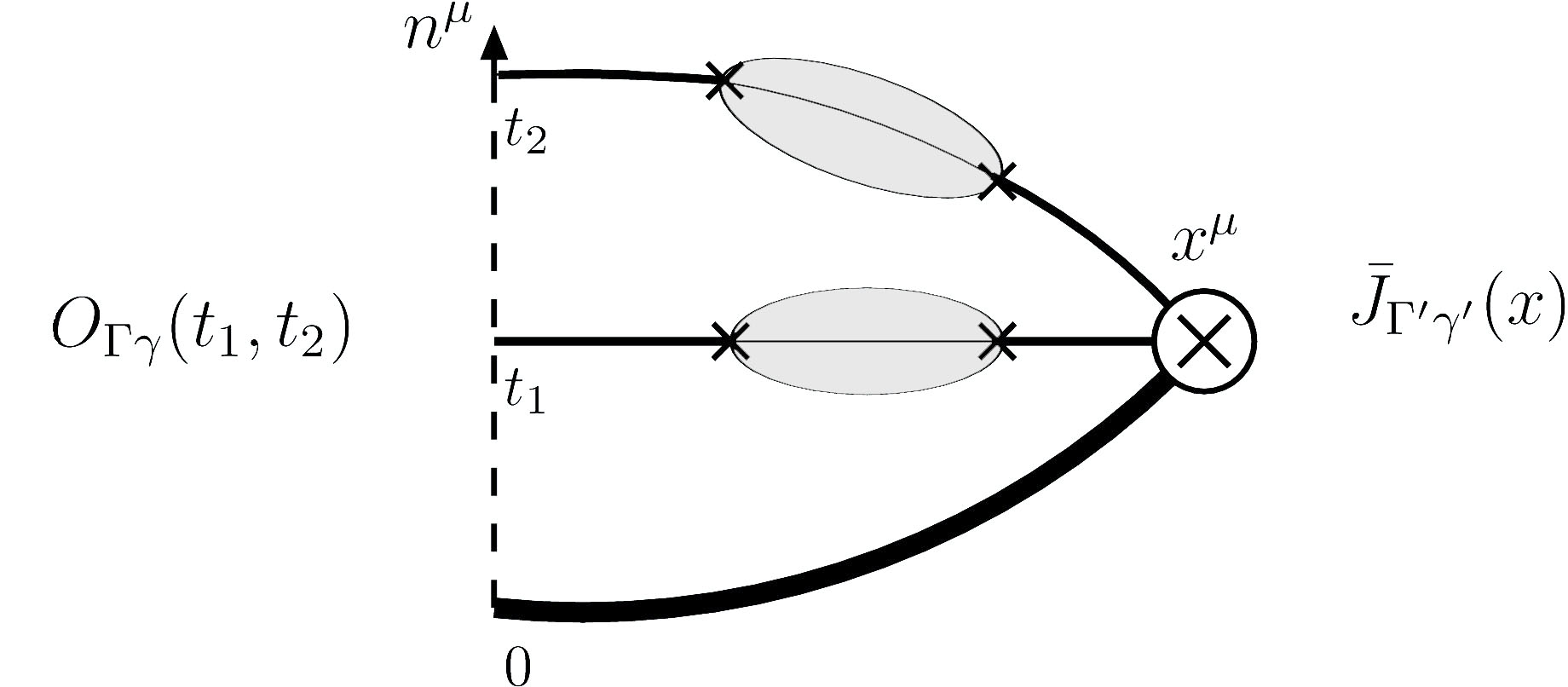}
}
\caption{
The correlation function in the QCD background. 
The light quarks move from the point~$x$ in the 
QCD background (shaded regions) to the light-like 
direction~$n^\mu$   
(dashed line). 
The $b$-quark is indicated by the thick black line.
}
\label{corrfunctionpicture}
\end{figure}

Following the standard procedure of QCD sum 
rules~\cite{Shifman:1978bx,Shifman:1978by},  
we calculate the matrix elements defined 
in~\eqref{parallelcurrents} and~\eqref{transversalcurrents}
by approximating the baryonic state $\ket{\baryongeneral{1}}$ 
by the local current defined in~\eqref{thelocalcurrentdefinition}, 
{\it i.\,e.}, $\ket{\baryongeneral{1}(x)} \approx J (x) \, \ket{0}$. 
The corresponding correlation function is defined by the matrix element 
$
\Pi_{\Gamma\Gamma'\gamma\gamma'} (t_1, t_2, x) = i \bra{0} 
 {\rm T} \left \{ \nonlocaloperator_{\Gamma \gamma} (t_1, t_2) 
\overline{J}_{\Gamma' \gamma'} (x)\right\} \ket{0}
$, 
in which $\nonlocaloperator_{\Gamma \gamma} (t_1, t_2)$ 
can be any non-local operator defined in Eqs.~\eqref{parallelcurrents} 
and \eqref{transversalcurrents}. The coefficient functions describe 
the propagation of the quarks inside the baryon from point~$x$ 
to the light-cone started from the point~$0$, which is 
parametrized by the light quark positions~$t_i$, presented  
in Fig.~\ref{corrfunctionpicture}. The procedure of constructing 
the QCD sum rules is well-known and results in the following general form:  
\begin{equation} 
\frac{1}{ 3}f_k \left ( 
A f^{(1)}_{\baryongeneral{1}} + B f^{(2)}_{\baryongeneral{1}} 
\right ) \widetilde\psi^{\rm SR}_k (t_1, t_2) \, 
{\rm e}^{- \bar\Lambda/\tau} = 
{\cal B} [ \Pi_k ] (t_1, t_2; \tau, s_0) , 
\label{eq:SRs-gen-PS}
\end{equation} 
where $f_k = f^{(1)}_{\baryongeneral{1}}$ for the even twists 
and $f_k = f^{(2)}_{\baryongeneral{1}}$ for the twist-3 
parallel distribution amplitudes (and swapped according 
to~\eqref{transversalcurrents} for the transverse twist). 
The effective baryon mass is introduced as the difference 
$\bar\Lambda = m_{\baryongeneral{1}} - m_b$,~$\tau$~is the Borel 
parameter, and~$s_0$ is the continuum threshold. 
The r.h.s. in Eq.~(\ref{eq:SRs-gen-PS}) is the Borel-transformed 
continuum-subtracted invariant function determined through 
the correlation function~$\Pi_k (t_1, t_2; \tau, s_0)$.  

The Fourier transform of the correlation function 
is then given by
\begin{align}
\Pi_{\Gamma\Gamma' \gamma\gamma'} (\omega_1, \omega_2; E) = 
\int\limits_{-\infty}^\infty \frac{d t_1}{2 \pi} \, e^{i \omega_1 t_1}
\int\limits_{-\infty}^\infty \frac{d t_2}{2 \pi} \, e^{i \omega_2 t_2} 
\int d^4x \, e^{-i E v.x} 
\bra{0} \nonlocaloperator_{\Gamma\gamma} (t_1, t_2) 
\overline{J}_{\Gamma' \gamma'} (x) \ket{0} .
\label{eq_corrsr}
\end{align}
Inserting and contracting the quark fields and 
performing the Fourier transformations yields
\begin{align} 
\Pi_{\Gamma\Gamma'} (\omega, u; E)_{\gamma\gamma'} = 
- 6i \int d^4 x \, e^{-i E v.x} \, 
\left [ \tilde{S}_{m=\infty} \right ]_{\gamma\gamma'} (x)
\trace{\Gamma \tilde S_{q_2} (\bar u \omega, t) 
       \Gamma'\tilde S_{q_1}^T  (u \omega, t)} .
\label{ersterstandvonpertpart}
\end{align}
The propagators~$\tilde{S}_q$, discussed later, are not 
free but describe the dynamical evolution of the valence 
quarks within the low-energetic QCD background of the 
baryon by including condensate contributions.

The heavy quark spin structure can be seen directly 
and is given by~$P_+$, see Eq.~\eqref{spinsumboth}.  
The light quark polarizations are given implicitly 
inside the trace by the~$\Gamma$ and~$\Gamma'$ structures. 
The correlation functions must be proportional to the 
Lorentz structure given by the spin sum~\eqref{spinsumboth}. 
The currents are then automatically projected on the proper 
polarization, such that
\begin{align}
\Pi_{\Gamma_{\perp\mu} \Gamma'_{\perp\nu} \gamma\gamma'} (\omega, u; E) 
&=
\Pi_\perp (\omega, u; E) 
\left ( - g_{\mu\nu} + v_\mu v_\nu + 
        \varepsilon_{\parallel\mu} \varepsilon_{\parallel\nu}/
        \varepsilon_\parallel^2
\right ) \left [ P_+ \right ]_{\gamma \gamma'} ,
\nonumber\\
\Pi_{\Gamma_{\parallel\mu} \Gamma'_{\parallel\nu} \gamma \gamma'} 
(\omega, u; E)  &=
- \Pi_\parallel (\omega, u; E) 
\left ( \varepsilon_{\parallel\mu} \varepsilon_{\parallel\nu}/ 
        \varepsilon_\parallel^2 \right ) 
\left [ P_+ \right ]_{\gamma\gamma'} ,
\end{align}
which defines the scalar functions $\Pi_\perp (\omega, u; E)$ 
and $\Pi_\parallel (\omega, u; E)$. They are identical for the 
parallel and transverse polarizations up to different 
contributions of the local interpolating currents, which gives 
\begin{equation}
\Pi_\perp (\omega, u; E)_{A, B} = \Pi_\parallel (\omega, u; E)_{B, A}.
\end{equation}
Since the LCDAs are given for the central value $A = B = 1/2$, 
the distribution functions $\psi (\omega, u: \mu)$ are 
the same for the transverse and parallel polarizations.
The dependence of the correlation functions on~$A$ and~$B$ 
is due to our choice of the set of linear independent operators 
for the non-local currents in~\eqref{parallelcurrents} 
and~\eqref{transversalcurrents}. Rewriting the parallel 
currents by introducing a different basis, 
for example~$\gamma_\parallel^\mu$, 
would change the above described symmetry in~$A$ and~$B$.

The correlation functions are evaluated 
in the following by using configuration space techniques, 
instead of working in the more common momentum space. 
The reason is that the leading order sunrise-diagrams 
become very simple in the coordinate space. 
Sunrise diagrams are certain types of $1 \to 1$ 
multi-loop self energy diagrams.
(For a review about the evaluation of diagrams of the 
sunrise type in coordinate space see~\cite{Groote:2005ay}).
The non-local diagrams, which are necessary for the calculation 
of the LCDAs, differ in principle from the sunrise type 
because they do not involve two local interpolating currents, 
instead they are ripped open along a light-like line at the left 
side of the diagram pictured in Fig.~\ref{corrfunctionpicture}. 
Hence, a more appropriate term would be lacerated sunrise diagram. 
But despite the difference, configuration space techniques 
work also well for this kind of diagrams (as shown below).

In the following we shortly introduce the different propagators 
of light and heavy quarks in the configuration space which are 
needed for the calculation of the correlation functions. In 
the heavy quark limit the Fourier transform of the heavy quark 
propagator has the very simple form of a classical point-like  
particle with a non-relativistic on-shell Dirac structure:
\begin{align}
\tilde S_{m=\infty} (x) &= 
\frac{1+ \sla{v}}{2} \int dt \, \delta^4 (x-vt).
\end{align}
To take the effects of the QCD background for the propagators 
of the light quark fields into account, the method of non-local 
condensates~\cite{Mikhailov:1986be,Mikhailov:1991pt}   is used. The propagator is then given by a sum of 
the free propagator $S_q (x)$ and the universal non-perturbative 
part $\mathcal{C}_q (x)$:
\begin{equation}
\begin{array}{lllll}
\raisebox{-1pt}[-20pt][-20pt]{$ \scriptstyle{\tilde{S}_q (x)}  $}
& &
\raisebox{-1pt}[-20pt][-20pt]{$ \scriptstyle{S_q (x)}          $}
& &
\raisebox{-1pt}[-20pt][-20pt]{$ \scriptstyle{\mathcal{C}_q (x)}$}
\\
\raisebox{-5pt}{\includegraphics[width=0.2\textwidth]{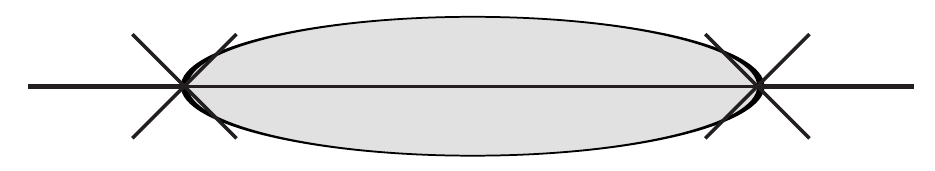}}
& = &
\raisebox{0pt}{\includegraphics[width=0.2\textwidth]{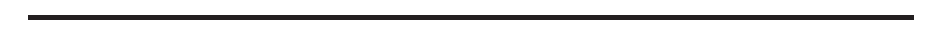}}
& + &
\raisebox{-5pt}{\includegraphics[width=0.2\textwidth]{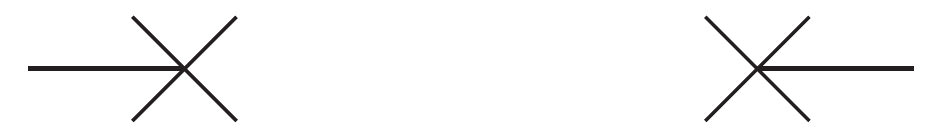}}
\end{array}
\label{nonfreepropexpansion}
\end{equation}
with
\index{Vacuum condensates non-local}
\begin{equation}
S_q (x) =
\frac{i}{2\pi^{2}} \, \frac{\sla{x}}{x^4} - \frac{m_q}{4 \pi^2 x^2},
\en\en\en\en\en\en
\mathcal{C}_q(x) =
\frac{1}{12} <\bar q (x) q (0)> ,
\label{eq_hqpropconfspace}
\end{equation}
\index{Propagator light quark in QCD background}
in which color and spin indices are omitted since the color structure 
is simply given by the Kronecker delta. For the non-local condensate 
we adopt the model proposed in~\cite{Braun:1994jq,Braun:2003wx}:
\begin{align}
& 
< \bar q (x) q (0) > = 
< \bar q q > \int\limits_0^\infty d\nu e^{\nu x^2/4} f (\nu) ,
\label{eq:nonlocalcondmodel} \\
& 
f (\nu) = \frac{\lambda^{a-2}}{\Gamma(a-2)} \, 
\nu^{1-a} \, e^{-\lambda/\nu} ,
\qquad
a-3 = \frac{4 \lambda}{m_0^2} ,
\nonumber 
\end{align}
in which $<\bar q q>$ is the local condensate,~$m_0^2$ 
is the ratio of the mixed quark-gluon and quark 
condensate, and~$\lambda$ is the correlation length. 
For completeness we also give the Fourier transformed 
operators ($t_i \to \omega_i$, $\omega_1 = u \omega$, 
$\omega_2 = \bar u \omega$), which are used in 
Eq.~\eqref{ersterstandvonpertpart}:
\begin{align} 
S_q (u \omega, t) &= - i e^{i \, \frac{u \omega t}{2 v_+}} \, 
\frac{2 m_q t v_+ + u \omega t v_+ \sla{\bar n} + 2 i \sla{n}}
     {8 \pi t^2 v_+^2},
\notag \\
\mathcal{C}_q (u \omega, t) &=
\frac{\pi <\bar q q>}{3 t v_+} \, 
e^{i \, \frac{t u \omega}{2 v_+}} \, 
f \left ( \frac{2 u \omega i}{t v_+} \right ) .
\label{eq_finallqprop}
\end{align}
The non-local condensate can be interpreted as the contribution 
of the quark, which descends to the non-perturbative sea of particles 
populating the QCD background, at one point and reemerges at 
a different point. It is still a question to what extent this 
sea is universal to the collective of all hadrons.
Moreover, little is known about the shape of these functions, 
and different models, such as~\cite{Bakulev:2002hk}, have been 
proposed in the literature to describe this non-perturbative behavior.

Inserting the propagators in the QCD background,  
Eq.~\eqref{ersterstandvonpertpart} reads diagrammatically 
\begin{equation}
\Pi  (\omega, u; E)  =
\raisebox{-14pt}{\includegraphics[width=0.1\textwidth]{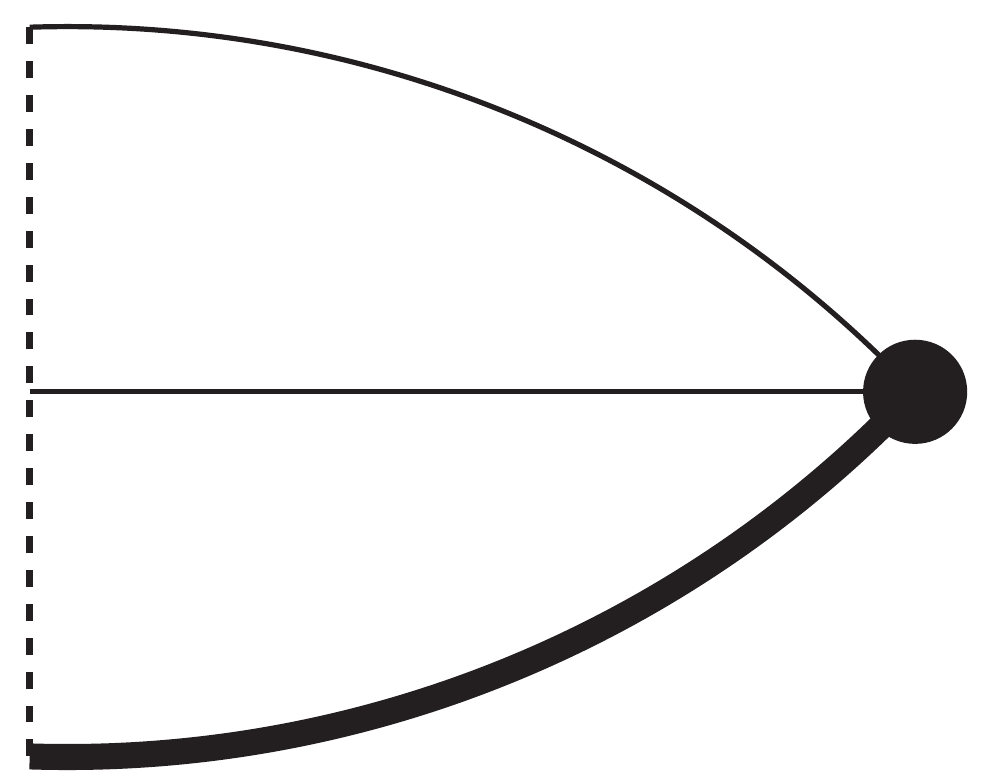}}+
\raisebox{-14pt}{\includegraphics[width=0.1\textwidth]{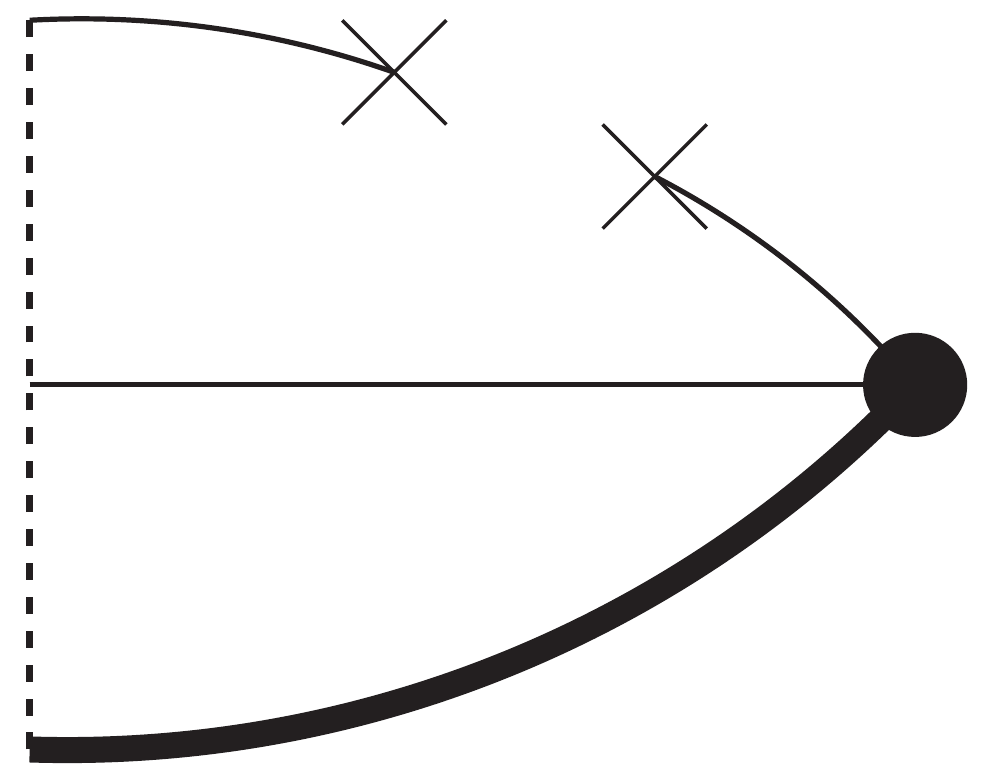}}+
\raisebox{-14pt}{\includegraphics[width=0.1\textwidth]{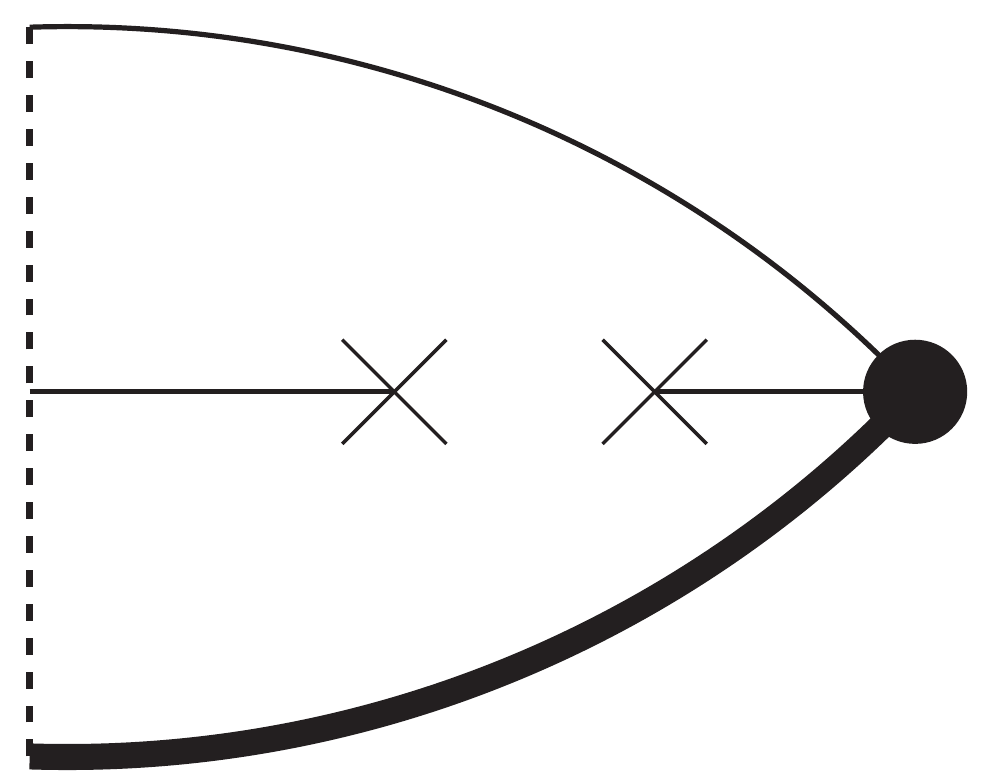}}+
\raisebox{-14pt}{\includegraphics[width=0.1\textwidth]{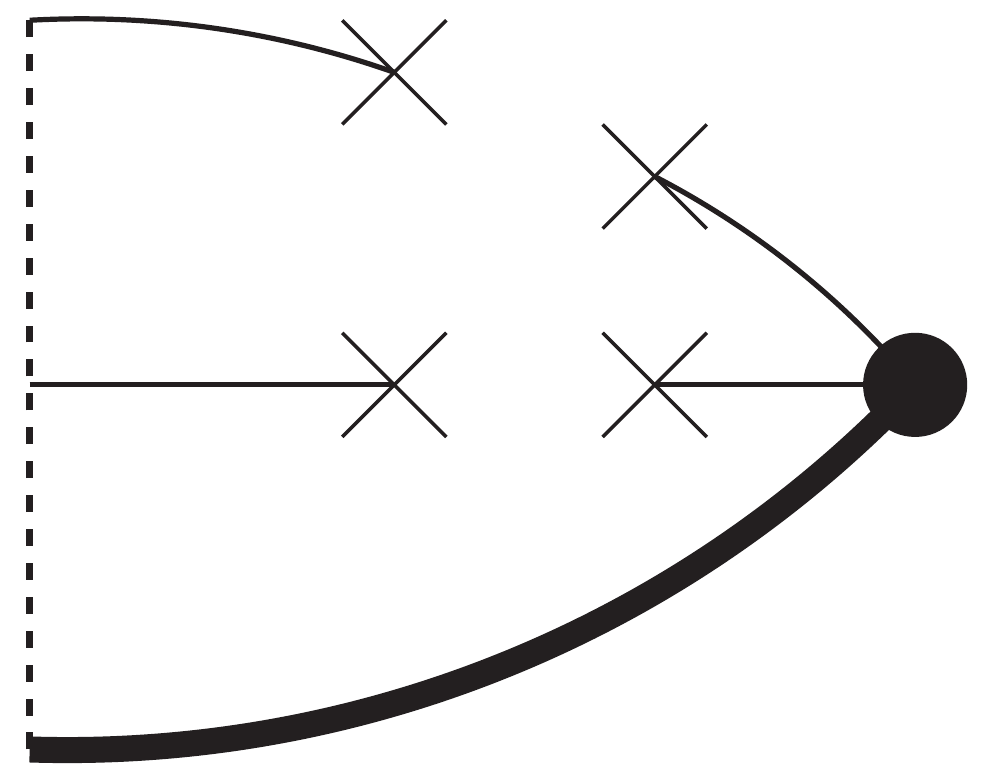}},
\label{eq_diaggrammaaticcorrellationfunction}
\end{equation}
in which the thin lines correspond to the light quarks 
and the thick line corresponds to the $b$-quark.
There is no $b$-quark condensate term, since it is suppressed 
by~$1/m_b$, and hence it vanishes in the heavy quark limit.

The Borel transform can easily be applied together 
with the Fourier transform, since for a function $f (t)$
\begin{equation}
\mathds{B} \left [ 
\int e^{i E t} \, f (t) \, dt 
\right ]_{E \to \tau} = i \, f \left ( \frac{i}{\tau} \right ) 
\label{eq:Borel-Fourier-trans} 
\end{equation}
holds. As known, the Borel transform handles the~UV divergences 
automatically, making renormalization rather easy. As a short reminder, 
the renormalization condition is applied by adding a certain polynomial 
in~$E$ to remove the UV divergent terms, which vanishes due to 
the derivative of arbitrary order in~$E$ in the Borel transform. 

The momentum cutoff is applied by cutting off the upper spectrum 
of the Laplace transform $L^{-1}$ (mapping to the spectral density) 
of the perturbative contributions, {\it i.\,e.},
\begin{equation}
f (\tau)
\stackrel{L^{-1}}{\longrightarrow}
\int\limits_{s_{\rm min}}^\infty \, 
e^{-s/\tau} \, \tilde f (s) \, ds
\stackrel{\tn{cutoff}}\longrightarrow
\int\limits_{s_{\rm min}}^{s_0} \, 
e^{-s/\tau} \, \tilde f (s) \, ds,
\label{eq:Laplace-transform}
\end{equation}
using
\begin{equation}
e^{-\Omega/\tau} \, \tau^a =
\frac{1}{\Gamma(a)}
\int\limits_\Omega^\infty d s \, e^{-s/\tau} \, (s-\Omega)^{a-1}.
\label{eq:Gamma-represent}
\end{equation}
One has to be careful, however, what one calls perturbative since
there might be mixed perturbative and condensate contributions.
Mixed contributions originate from diagrams in which one light quark 
is described by the low-energy condensate and the other light quark 
is described by the high-energy perturbative propagator.
We adopted the procedure of~\cite{Ball:2008fw} and call 
the distribution perturbative when at least one perturbative 
term is present. This procedure leads to incomplete~$\Gamma$-functions. 
For book keeping, we introduce the function
\begin{equation}
E_a (x) \equiv \frac{1}{\Gamma (a + 1)} 
\int_0^x dt \, t^a {\rm e}^{-t} =
1 - \frac{\Gamma (a + 1, x)}{\Gamma(a + 1)} ,
\label{eq:Ea-func-def}
\end{equation}
in which $\Gamma (a + 1, x) = \int_x^\infty dt \, t^a {\rm e}^{-t}$
is the incomplete $\Gamma$-function. 
The application of the quark-hadron duality thus results in
$e^{-\Omega/\tau}\, \tau^a \to e^{-\Omega/\tau} \, \tau^a \, 
E_{a-1} (s_0 - \Omega)$. For $a=N \in \mathds{N}$, 
this function is reduced to the well-known form
\begin{equation}
E_N (x) = 1 - {\rm e}^{- x} \sum_{n = 0}^N \frac{x^n}{n !} .
\label{eq:EN-func-def}
\end{equation}
In numerical estimates, it is also helpful to use the relation
\begin{equation}
E_a (x) = E_{a+1} (x)
+ \frac{x^{a + 1} \, {\rm e}^{-x}}{\Gamma(a + 2)} ,
\label{eq:rising-procedure}
\end{equation}
obtained from~\eqref{eq:Ea-func-def} after integration by parts, 
connecting the negative value of the parameter~$a$ in~$E_a (x)$
with a corresponding positive one.

Inserting the propagators~\eqref{eq_hqpropconfspace} 
and~\eqref{eq_finallqprop} in Eq.~\eqref{ersterstandvonpertpart} 
and performing the Borel and Fourier transforms, the sum rules 
defined in~\eqref{eq:SRs-gen-PS} are obtained in a straightforward way.
We summarize our sum rule results where the leading-twist transverse 
formula is given as
\begin{eqnarray}
&&
\frac{1}{ 3}f^{(2)}_{\baryongeneral{1}} \left (
A f^{(1)}_{\baryongeneral{1}} + B f^{(2)}_{\baryongeneral{1}}
\right ) \tilde \psi_\perp^n (\omega, u) \,
{\rm e}^{- \bar\Lambda/\tau} =
\label{eq:psi2-SR-total} \\ 
&& \quad \hphantom{=}
\frac{3 \tau^4}{2 \pi^4} \,
\left [ B \hat \omega^2 \, u \bar u
+ A \, \hat \omega \left ( \hat m_2 u + \hat m_1 \bar u \right )
\right ] E_1 ( 2 \hat s_\omega )
{\rm e}^{- \hat \omega}
\nonumber \\ 
&& \quad -
\frac{\langle \bar q_1 q_1 \rangle \tau^3}{\pi^2} \,
\left [ A \hat \omega \bar u + B \hat m_2 \right ]
f (2 \tau \omega u) \, E_{2 - a} ( 2 \hat s_\kappa ) \,
{\rm e}^{- \hat \omega}
\nonumber \\ 
&& \quad -
\frac{\langle \bar q_2 q_2 \rangle \tau^3}{\pi^2} \,
\left [ A \hat \omega u + B \hat m_1 \right ]
f (2 \tau \omega \bar u) \, E_{2 - a} ( 2 \hat s_{\bar\kappa} ) \,
{\rm e}^{- \hat \omega}
\nonumber \\ 
&& \quad + \frac{2 B}{3} \,
\langle \bar q_1 q_1 \rangle \, \langle \bar q_2 q_2 \rangle \,
\tau^2 \, f (2 \tau \omega u) \, f (2 \tau \omega \bar u) \,
E_{3 - 2 a} ( 2 \hat s_{\kappa \bar\kappa} ) \,
{\rm e}^{- \hat \omega} ,
\nonumber 
\end{eqnarray}
in which 
$s_\omega = s_0 - \omega/2$,
$\hat m = m/(2\tau)$,
$\hat\omega = \omega/(2\tau)$,
$\hat s_\omega = s_\omega/(2\tau)$,
$\hat s_\kappa = \hat s_\omega - \kappa/2$,
$\hat s_{\bar\kappa} = \hat s_\omega - \bar\kappa/2$,
$\hat s_{\kappa \bar\kappa} = \hat s_\omega - \kappa/2 - \bar\kappa/2$,
and the short-hand notations
\begin{equation}
\kappa = \frac{\lambda}{2 u \omega \tau},
\qquad
\bar\kappa = \frac{\lambda}{2 \bar u \omega \tau} 
\label{eq:kappa-def}
\end{equation}
are used.
For the twist-3 and twist-4 LCDAs, the results can be derived analogously: 
\begin{eqnarray}
&&
\frac{1}{ 3}f^{(1)}_{\baryongeneral{1}} \left (
A f^{(1)}_{\baryongeneral{1}} + B f^{(2)}_{\baryongeneral{1}}
\right ) \tilde \psi_\perp^\Id (\omega, u) \,
{\rm e}^{- \bar\Lambda/\tau} 
\label{eq:psi3s-SR-total} \\ 
&& \quad = 
\frac{3 \tau^4}{4 \pi^4} \left \{
\left [ A \hat \omega + B \, (\hat m_1 + \hat m_2) \right ]
E_2 (2 \hat s_\omega)
+ B \hat \omega \, ( \hat m_2 u + \hat m_1 \bar u ) \,
E_1 (2 \hat s_\omega)
\right \} {\rm e}^{- \hat\omega}
\nonumber \\ 
&& \quad -
\frac{\langle \bar q_1 q_1 \rangle \tau^3}{2 \pi^2}
\left [ B \, E_{3 - a} (2 \hat s_\kappa)
+ \left ( B \hat \omega \bar u + 2 A \hat m_2 \right )
E_{2 - a} (2 \hat s_\kappa) \right ]
f (2 \tau \omega u) \, {\rm e}^{- \hat\omega}
\nonumber \\ 
&& \quad -
\frac{\langle \bar q_2 q_2 \rangle \tau^3}{2 \pi^2}
\left [ B \, E_{3 - a} (2 \hat s_{\bar\kappa})
+ \left ( B \hat \omega u + 2 A \hat m_1 \right )
E_{2 - a} (2 \hat s_{\bar\kappa}) \right ]
f (2 \tau \omega \bar u) \, {\rm e}^{- \hat\omega}
\nonumber \\ 
&& \quad + \frac{2 A}{3} \,
\langle \bar q_1 q_1 \rangle \, \langle \bar q_2 q_2 \rangle \,
\tau^2 \, f (2 \tau \omega u) \, f (2 \tau \omega \bar u) \,
E_{3 - 2 a} (2 \hat s_{\kappa \bar\kappa}) \,
{\rm e}^{- \hat\omega} ,
\nonumber \\
\nonumber \\ 
&& 
\frac{1}{ 3}f^{(1)}_{\baryongeneral{1}} \left (
A f^{(1)}_{\baryongeneral{1}} + B f^{(2)}_{\baryongeneral{1}}
\right ) \tilde \psi_\perp^{n\bar{n}} (\omega, u) \,
{\rm e}^{- \bar\Lambda/\tau} 
\label{eq:psi3sig-SR-total} \\ 
&& \quad = 
\frac{3 \tau^4}{4 \pi^4} \left \{ \left [
A \hat\omega  \, (u - \bar u) + B \, (\hat m_1 - \hat m_2) \right ]
E_2 (2 \hat s_\omega)
+ B \hat\omega \, ( \hat m_2 u - \hat m_1 \bar u )
E_1 (2 \hat s_\omega)
\right \} {\rm e}^{- \hat\omega}
\nonumber \\ 
&& \quad -
\frac{B \langle \bar q_1 q_1 \rangle \tau^3}{2 \pi^2} \,
\left [ E_{3 - a} (2 \hat s_\kappa)
- \hat\omega \bar u \, E_{2 - a} (2 \hat s_\kappa) \right ]
f (2 \tau \omega u) \, {\rm e}^{- \hat\omega}
\nonumber \\ 
&& \quad +
\frac{B \langle \bar q_2 q_2 \rangle \tau^3}{2 \pi^2} \,
\left [ E_{3 - a} (2 \hat s_{\bar\kappa})
- \hat\omega u \, E_{2 - a} (2 \hat s_{\bar\kappa}) \right ]
f (2 \tau \omega \bar u) \, {\rm e}^{- \hat\omega} ,
\nonumber \\ 
\nonumber \\  
&& 
\frac{1}{ 3}f^{(2)}_{\baryongeneral{1}} \left (
A f^{(1)}_{\baryongeneral{1}} + B f^{(2)}_{\baryongeneral{1}}
\right ) \tilde \psi_\perp^{\bar{n}}(\omega, u) \,
{\rm e}^{- \bar\Lambda/\tau} 
\label{eq:psi4-SR-total} \\ 
&& \quad = 
\frac{3 \tau^4}{2 \pi^4}
\left [ B \, E_3 (2 \hat s_\omega)
+ A \left ( \hat m_1 + \hat m_2 \right )
E_2 (2 \hat s_\omega)
\right ] {\rm e}^{- \hat\omega}
\nonumber \\ 
&& \quad -
\frac{\langle \bar q_1 q_1 \rangle \tau^3}{\pi^2} \,
\left [ A \, E_{3 - a} (2 \hat s_\kappa)
+ B \hat m_2 \, E_{2 - a} (2 \hat s_\kappa)
\right ] f (2 \tau \omega u) \,
{\rm e}^{- \hat\omega}
\nonumber \\ 
&& \quad -
\frac{\langle \bar q_2 q_2 \rangle \tau^3}{\pi^2} \,
\left [ A \, E_{3 - a} (2 \hat s_{\bar\kappa})
+ B \hat m_1 \, E_{2 - a} (2 \hat s_{\bar\kappa})
\right ] f (2 \tau \omega \bar u) \,
{\rm e}^{- \hat\omega}
\nonumber \\ 
&& \quad + \frac{2 B}{3} \,
\langle \bar q_1 q_1 \rangle \, \langle \bar q_2 q_2 \rangle \,
\tau^2 \, f (2 \tau \omega u) \, f (2 \tau \omega \bar u) \,
E_{3 - 2 a} (2 \hat s_{\kappa \bar\kappa}) \,
{\rm e}^{- \hat\omega}.
\nonumber 
\end{eqnarray}
The result for the parallel counterpart can be obtained 
by the replacement $A \leftrightarrow B$.

The QCD sum rules given in~\eqref{eq:psi2-SR-total} 
can not be used in calculations directly. 
The main reason for this is that the sum rules are built 
from a patchwork of different contributions, the perturbative 
and condensate parts. They show neither smooth behavior,
nor necessarily the correct asymptotic behavior, 
{\it i.\,e.}, the asymptotic behavior of the perturbative 
contribution. As a consequence, one has to propose model 
functions which are then constrained by the sum rules.
The LCDA models are discussed below in Sec.~\ref{lcdamodel}.

\section{LCDA Models} 
\label{lcdamodel}
\index{Model $b$-baryon LCDAs}
\begin{figure}[t]
\centering
\resizebox{0.5\textwidth}{0.32\textwidth}{
\includegraphics[width=0.5\textwidth]{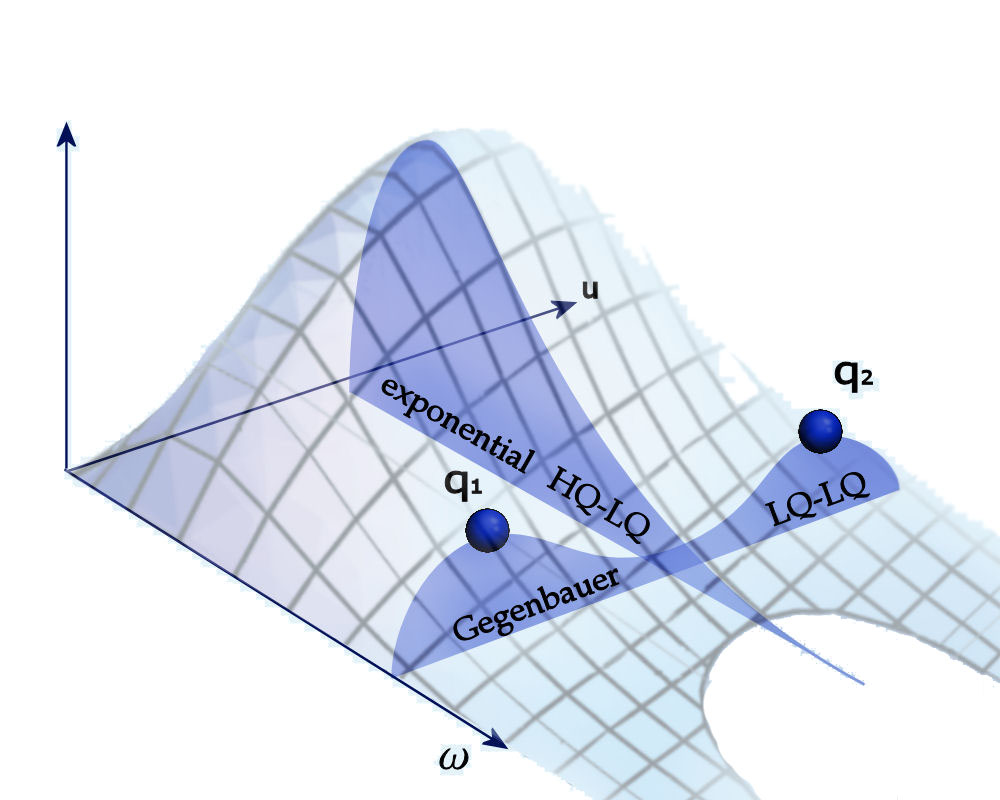}
}
\caption{
Model functions for the $b$-baryon LCDAs, composed 
of the exponential part for the heavy-light interaction 
and the Gegenbauer polynomials for the light-light interaction.
}
\label{fig_gegenexpo}
\end{figure}
\begin{table}[t]
\caption{
Moments for the LCDA model functions 
in terms of their parameters.}  
\label{momentsmodel}
\begin{center}
\begin{tabular}{c|cccccc}
\hline\hline
twist                                   &
$\langle 1 \rangle$                     &
$\langle \omega^{-1} \rangle$           &
$\langle C^{3/2}_1 \rangle$             &
$\langle \omega^{-1} C^{3/2}_1 \rangle$ &
$\langle C^{3/2}_2 \rangle$             &
$\langle \omega^{-1} C^{3/2}_2 \rangle$ \\ \hline
$ 2     $ & $a_0       $ & $a_0/3  \varepsilon_0      $ & $ 3 a_1/5 $ & $   a_1/5  \varepsilon_1$ & $ 3 a_2/7 $ & $  a_2/7  \varepsilon_2  $  \\
\hline
twist                                   &
$\langle 1 \rangle$                     &
$\langle \omega^{-1} \rangle$           &
$\langle C^{1/2}_1 \rangle$             &
$\langle \omega^{-1} C^{1/2}_1 \rangle$ &
$\langle C^{1/2}_2 \rangle$             &
$\langle \omega^{-1} C^{1/2}_2 \rangle$ \\ \hline
$ 3     $ & $a_0      $ & $a_0/2  \varepsilon_0      $ & $ a_1 $ & $  a_1/2  \varepsilon_1$ & $ a_2 $ & $  a_2/2  \varepsilon_2  $  \\
$ 4     $ & $a_0      $ & $a_0/ \varepsilon_0        $ & $ a_1 $ & $  a_1/ \varepsilon_1  $ & $ a_2 $ & $  a_2/ \varepsilon_2    $  \\
\hline\hline
\end{tabular}
\end{center}
\end{table}
In the parametrization, in which the baryon is described 
by the total energy of the light quark system~$\omega$ and 
the momentum fractions~$u$ and~$\bar u$, the following arguments become manifest. 
The dynamics of the heavy-light baryon part, {\it i.\,e.}, the dynamics of the heavy quark and the light diquark, is described with the same 
formalism which has worked well for the heavy-light meson dynamics. 
The dynamics of the light quarks in the diquark is then described 
in the same way as the light-light mesons. This ansatz is consistent 
with the equations of motion for the light quarks~\cite{Ball:1998sk}.
In conclusion we choose the multiplicative ansatz from~\cite{Ball:2008fw}. 
In this approach the Gegenbauer polynomials 
$C^\lambda_n (x)$~\cite{Bateman-vII} 
take the light quarks into account and an exponential factor characterizes 
the dynamics of the heavy-light system, as shown in Fig.~\ref{fig_gegenexpo}. 
\index{Gegenbauer polynomials}
It is more convenient to choose $C^{3/2}_n (x)$ Gegenbauer polynomials 
for twist-2 but $C^{1/2}_n (x)$ for the other twist functions 
similar to the expansion of the vector mesons~\cite{Ball:1998sk}.  
The first three polynomials are sufficient to account 
for the precision in this work. They are defined as follows:
\begin{equation}
C^\lambda_0 (x) = 1 , 
\enspace\enspace\enspace
C^\lambda_1 (x) = 2 \lambda x, 
\enspace\enspace\enspace
C^\lambda_2 (x) = 2 \lambda \left ( 1 + \lambda \right ) x^2 - \lambda .
\label{eq:Gegenbauer-polinomials}  
\end{equation}
To obtain the model fit, we calculate the momentum fraction 
integrals (the moments) which are defined for an arbitrary 
function $f (\omega, u)$ as:
\begin{equation}
\langle f (\omega, u) \rangle^{\baryongeneralW}_k \equiv
\int_0^{2 s_0} \omega d\omega \int_0^1 du \,
f (\omega, u) \, \tilde \psi^{\rm SR}_k (\omega, u) .
\label{eq:moments-SR-def}
\end{equation}
The model functions for the LCDAs of different twists are:
\begin{eqnarray}
\tilde \psi_2 (\omega, u) & = &
\omega^2 u (1 - u) \, \sum_{n = 0}^2
\frac{a_n}{{\epsilon_n}^4} \, 
\frac{C^{3/2}_n (2 u - 1)}{| C^{3/2}_n |^2} \,
{\rm e}^{- \omega/\epsilon_n} ,
\label{eq:LCDA2-model-def} \\ 
\tilde \psi_3 (\omega, u) & = &
\frac{\omega}{2} \, \sum_{n = 0}^2
\frac{a_n}{{\epsilon_n}^3} \, 
\frac{C^{1/2}_n (2 u - 1)}{| C^{1/2}_n |^2} \,
{\rm e}^{- \omega/\epsilon_n} ,
\label{eq:LCDA3s-model-def} \\ 
\tilde \psi_4 (\omega, u) & = &
\sum_{n = 0}^2
\frac{a_n}{{\epsilon_n}^2} \, 
\frac{C^{1/2}_n (2 u - 1)}{| C^{1/2}_n |^2} \,
{\rm e}^{- \omega/\epsilon_n} ,
\label{eq:LCDA4-model-def}
\end{eqnarray}
in which the twist is indicated by the subscript numbers and 
\begin{equation} 
\left | C^\lambda_n \right |^2 = 
\int_0^1 du \left [ C^\lambda_n (2 u - 1) \right ]^2 , 
\label{eq:Gegenbauer-norma} 
\end{equation} 
with
$\big | C^{1/2}_0 \big |^2 = \big | C^{3/2}_0 \big |^2 = 1$,
$\big | C^{1/2}_1 \big |^2 = 1/3$, $\big | C^{3/2}_1 \big |^2 = 3$,
$\big | C^{1/2}_2 \big |^2 = 1/5$, and $\big | C^{3/2}_2 \big |^2 = 6$. 
The prefactors in front of the sums in Eq.~\eqref{eq:LCDA2-model-def}~---~\eqref{eq:LCDA4-model-def}
($\omega^2 u (1 - u)$, etc.) are determined by the corresponding
perturbative part in order to give the correct asymptotic behavior 
(compare the~$B$ term in the first line in Eq.~\eqref{eq:psi2-SR-total}).
The parameters~$\epsilon_i$ are strictly positive 
to satisfy the asymptotic behavior.
The moments of the functions~\eqref{eq:LCDA2-model-def}~---~\eqref{eq:LCDA4-model-def}
which are calculated with the use of~\eqref{eq:moments-SR-def} 
are listed in Tab.~\ref{momentsmodel}.

In the construction of the models for the LCDAs, we have 
truncated the Gegenbauer expansion at the second non-asymptotic 
term and have taken the limit $s_0 \to \infty$ in the integral 
over~$\omega$ which has negligible effect. 

\section{Results and Numerical Analysis} 
\label{sec_resultsLCDAs}
\begin{figure}[t]
\includegraphics[width=1\textwidth]{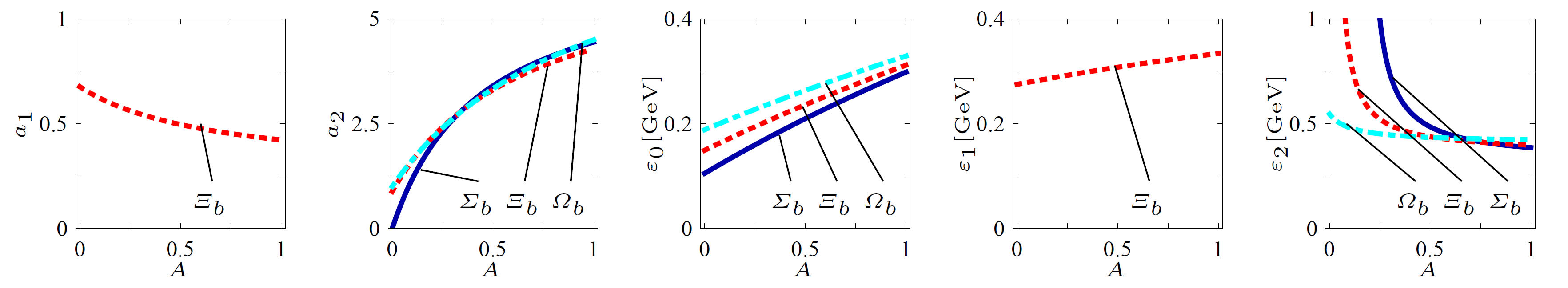}
\caption{
The dependence of the transverse twist-2 model parameters 
listed in Tab.~\ref{modelparam} on the variation of the 
coefficient~$A$ of the local interpolating currents (the 
parallel ones are obtained by the replacement $A \to 1 - A$). 
} 
\label{tw2modpar}
\end{figure} 

\begin{table}[tb]
\caption{
Input values from 
Refs.~\cite{Ball:2008fw,Shifman:1978by,Chetyrkin:2007vm,Chernyak:1983ej}.
For~$\bar\Lambda$ and~$s_0$, see Tab.~\ref{tab:baryon-masses}.
} 
\label{tab:NP-input} 
\begin{center} 
\begin{tabular}{lcc||clc} 
\\  
$m_{u,d}$ & 0  
&&&
$\langle \bar q q \rangle$   & $-242$~MeV$^3$ 
\\ 
$m_s$    & $128$~MeV 
&&& 
$\langle \bar s s \rangle/\langle \bar q q \rangle$ & $0.8$ 
\\ 
$\lambda$ & $0.16$~GeV$^2$ 
&&& 
$m_0^2 = \langle \bar q D^2 q \rangle/\langle \bar q q \rangle$ & $0.8$~GeV$^2$
\\ 
$\tau$ & $0.6$~GeV 
&&& 
$m_{s0}^2 = \langle \bar s D^2 s \rangle/\langle \bar s s \rangle$ & $1.36$~GeV$^2$
\end{tabular} 
\end{center} 
\end{table}

The moments of the functions defined 
in Eqs.~\eqref{eq:LCDA2-model-def}~---~\eqref{eq:LCDA4-model-def} can be calculated 
using~\eqref{eq:moments-SR-def}.
To perform a numerical analysis, we discuss and specify 
the required input parameters which are summarized in 
Tab.~\ref{tab:NP-input}. 
The values of the effective baryon masses 
$\bar\Lambda = m_{H_b^j} - m_b$ in the HQET for $m_b = 4.8$~GeV 
are presented in Tab.~\ref{tab:baryon-masses} where experimental 
measurements~\cite{Beringer:1900zz} and theoretical predictions 
(based on the HQET~\cite{Liu:2007fg} and Lattice QCD~\cite{Lewis:2008fu}) 
for the masses (in units of MeV) of the ground-state bottom 
baryons are also shown. A comparative analysis 
of predictions for the heavy-baryon masses can 
be found in Refs.~\cite{Liu:2007fg,Zhang:2008pm}. 
The continuum threshold values~$s_0$ (the last column 
in Tab.~\ref{tab:baryon-masses}) used by us are 
in agreement with the ones from~\cite{Liu:2007fg} derived for 
the baryon-mass evaluated to order~$1/m_b$ within the HQET. 
For the discussion of these parameters 
see~\cite{Chetyrkin:2007vm} and references therein. 
Note that the shape function~$f (\nu)$ in the non-local 
quark condensate~\eqref{eq:nonlocalcondmodel} is assumed 
to be flavor independent for all light quarks.

\begin{table}[t]
\caption{
Parameters for the model functions in 
Eqs.~\eqref{eq:LCDA2-model-def}~---~\eqref{eq:LCDA4-model-def} for the transverse LCDAs  
(the parallel ones are obtained by the replacement $A \to 1-A$ 
and the~$\Lambda_b$ and~$\Xi_b$ currents correspond 
to the parallel~$\Sigma_b$ and~$\Xi'_b$ ones, respectively). 
For twist-2 LCDAs, the parameter dependence on~$A$ is plotted 
in Fig.~\ref{tw2modpar}.  The twist notations 3a and 3s correspond to the LCDAs
$\psi^{n \bar n}$ and $\psi^\Id$, respectively.
}   
\label{modelparam}
\begin{center}
\def\arraystretch{1.5}
\begin{tabular}{p{0.6cm}|p{0.5cm}|p{1.3cm}p{1.3cm}p{1.3cm}p{1.3cm}p{1.3cm}p{1.3cm}}
\hline\hline \multirow{6}{*}{\mbox{\Large $ \Sigma_b $}}    
&twist& $a_0$ & $a_1$ & $a_2$ & $\varepsilon_0$[GeV] & $\varepsilon_1$[GeV] & $\varepsilon_2$[GeV]  \\ 
\cline{2-8} 
&$2      $  & $  1 $ & $ -  $ & $ \frac{6.4 A}{A+0.44}  $& $  \frac{1.4 A+0.6}{A+5.7} $& $ -  $ &$ \frac{0.32 A}{A-0.17}  $\\
&$3 s    $  & $  1 $ & $ -  $ & $ \frac{0.12 A-0.08}{A-1.4}  $& $ \frac{0.56 A-0.77}{A-2.6}  $& $ -  $ &$ \frac{0.25 A-0.16}{A+0.41}  $\\
&$3a     $  & $ -  $ & $ 1  $ & $ -  $& $ -  $& $ \frac{  0.35 A -0.43   }{A-1.2 }  $ &$ -  $\\
&$4      $  & $ 1  $ & $ -  $ & $ \frac{  -0.07 A - 0.05   }{A +0.34 }  $& $ \frac{  0.65 A+0.22   }{A+1 }  $& $ -  $ &$ \frac{  5.5 A+3.8   }{A +29 }  $\\
 \hline\hline \multirow{6}{*}{\mbox{\Large$ \Xi_b^{ \prime}  $}}  
&twist& $a_0$ & $a_1$ & $a_2$ & $\varepsilon_0$ & $\varepsilon_1$ & $\varepsilon_2$  \\ 
\cline{2-8}
&$2      $  & $1$ & $\frac{0.25 A+0.46}{A+0.68}$ & $\frac{6.6A+0.6}{A+0.68}$& $\frac{1.4 A+1}{A+6.7}$& $\frac{0.57A+1.1}{A+4}$ &$\frac{0.36 A+0.03}{A-0.02}$\\
&$3 s    $  & $1$ & $\frac{0.04A-0.14}{A-1.6}$ & $\frac{0.12 A-0.09}{A-1.6}$& $\frac{0.56 A-0.91}{A-2.9}$& $\frac{-27 A+92}{160}$ &$\frac{  0.3 A-0.24}{A+0.54}$\\
&$3a     $  & $\frac{-0.16 A+0.16}{A-1.3}$ & $1$ & $\frac{0.17 A-0.17}{A-1.3}$& $\frac{0.11A-0.11}{A-1}$& $\frac{0.39A-0.49}{A-1.3}$ &$\frac{0.33A-0.33}{A-1}$\\
&$4      $  & $1$ & $\frac{0.03 A+0.11}{A+0.16}$ & $\frac{-0.1A-0.03}{A+0.61}$& $\frac{0.63A+0.38}{A+1.3}$& $\frac{-0.82A-3.1}{A-3.9}$ &$\frac{1.2A+0.34}{A+4.1}$\\
 \hline\hline \multirow{6}{*}{\mbox{\Large $\Omega_b$}}  
&twist& $a_0$ & $a_1$ & $a_2$ & $\varepsilon_0$ & $\varepsilon_1$ & $\varepsilon_2$  \\ 
\cline{2-8}
&$2      $  & $1$ & $-$ & $\frac{8 A+1}{A+1}$& $\frac{1.3A+1.3}{A+6.9}$& $-$ &$\frac{0.41A+0.06}{A+0.11}$\\
&$3 s    $  & $1$ & $-$ & $\frac{0.17 A-0.16}{A-2}$& $\frac{0.56 A-1.1}{A-3.22}$& $-$ &$\frac{0.44A-0.43}{A+0.27}$\\
&$3a     $  & $-$ & $1$ & $-$& $-$& $\frac{0.45A-0.63}{A-1.4}$ &$-$\\
&$4      $  & $1$ & $-$ & $\frac{-0.10A-0.01}{A+1}$& $\frac{0.62A+0.62}{A+1.62}$& $-$ &$\frac{0.87A+0.07}{A+2.53}$\\
 \hline\hline
\end{tabular}
\end{center}
\end{table}

The calculation is performed in the $b$-baryon rest frame 
with $v_+ = 1$ at an energy scale of $\mu=1$~GeV.
The method of the non-local condensates which involves the 
parameters~$\lambda$ and~$m_0^2$ is not yet completely understood.
Especially, since there is only one model parameter known, 
namely the ratio 
\begin{align}
m_0^2 & = 
\frac{\langle \bar q D^2 q \rangle}{\langle \bar q q \rangle}
\label{eq:m0-def}
\end{align}
of the 5-dimensional and 3-dimensional local condensates. 
This parameter determines the center of the quark virtuality 
distribution in the QCD background, but is not sufficient 
to determine the shape of the quark distribution. To determine 
the form, also yet unknown dimension-7 local condensates are needed.
We took the shape parameter~$\lambda$ as the universal parameter 
which is not influenced by either the baryon or the mass of the 
propagating quark. For the strange quark, where 
\begin{align}
m_{s0}^2 = 
\frac{\langle \bar s D^2 s \rangle}{\langle \bar s s \rangle} ,
\label{eq:m0s-def}
\end{align}
the situation is more difficult since even dimension-5 
condensates are not yet clearly understood.
The value~$R$, which is defined by
\begin{align}
R = \frac{\langle \bar s D^2 s \rangle}{\langle \bar q D^2 q \rangle} ,
\label{eq:R-ratio}
\end{align}
varies from $R \approx 0.8$~\cite{Chernyak:1983ej,Lee:1997ix}
 which gives $m_0^2 \approx m_{s0}^2$
to values around $R \approx 1.3$~%
\cite{Beneke:1992ba,Dosch:2000ej,Dosch:1988vv} 
 which gives 
\begin{align}
m_{s0}^2 \approx 1.7 \, m_0^2,
\end{align}
which we took for our calculation since in this case the 
$SU(3)_{\rm F}$-breaking effects appear already in the lower 
$\omega$-spectrum. 
The influence of the choice of~$R$ on our results is within the  
already large uncertainties due to the other parameters.
%
%
%
\begin{figure}[t] 
\includegraphics[width=1\textwidth]{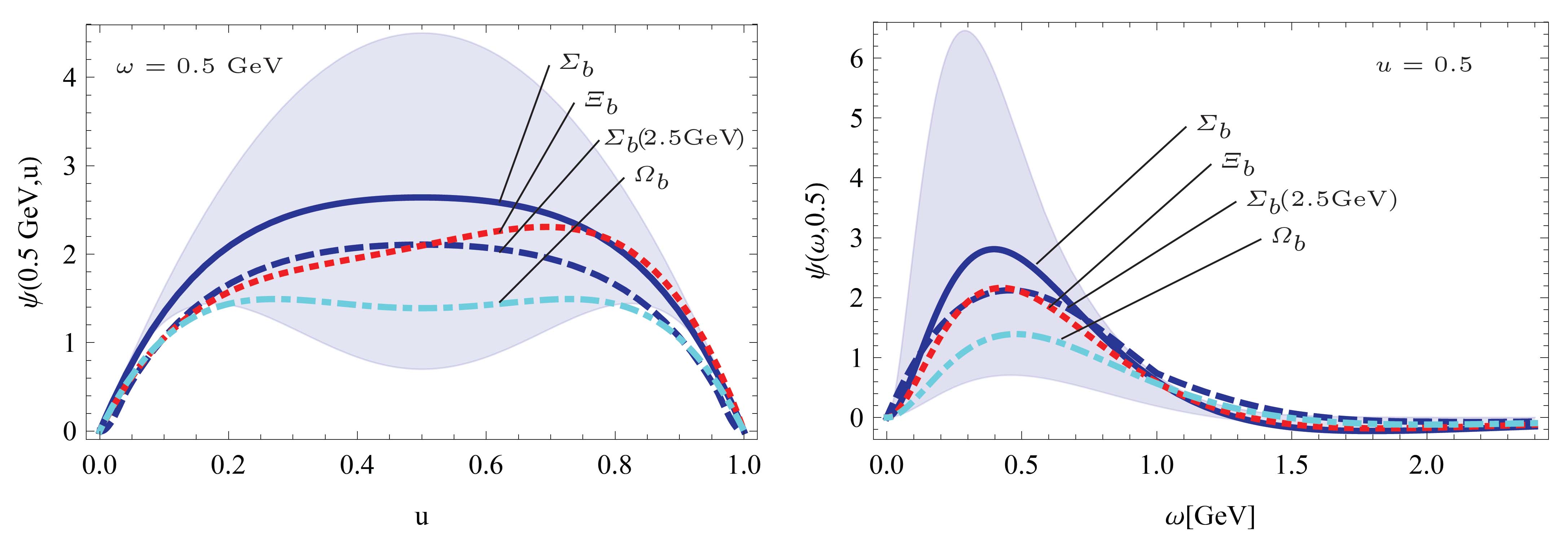}
\caption{(Color online)
 The Twist 2 functions of $\Sigma_b$ (blue, solid), $\Xi_b$ (red, dashed) and $\Omega_b$ (cyan, dot-dashed)
at the energy scale $\mu=1\; \tn{GeV}$ (blue, solid line) and the energy scale $\mu=2.5\; \tn{GeV}$
(blue, dashed line) including the most conservative error $A\in [0,1]$ for $\Sigma_b$ (blue, shaded region).
\label{alltwist2onevariable}}
\end{figure}
\begin{figure}[t]
\centering\hspace{0.4cm}
\includegraphics[width=7cm]{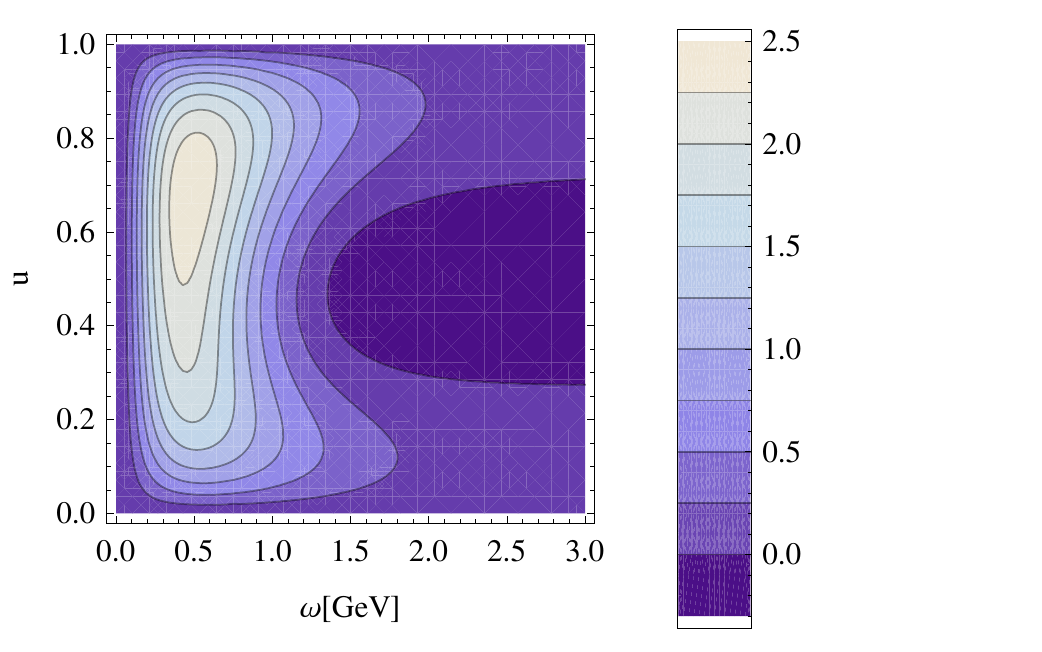}\hspace{-1.2cm}
\includegraphics[width=7cm]{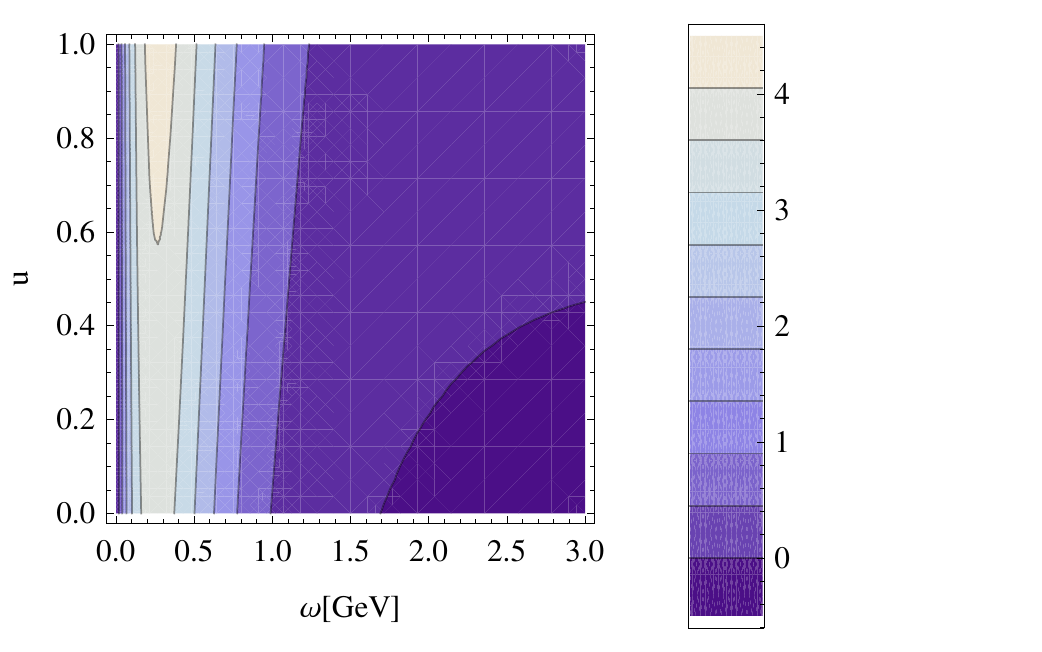}   
\\
\centering\hspace{0.4cm}
\includegraphics[width=7cm]{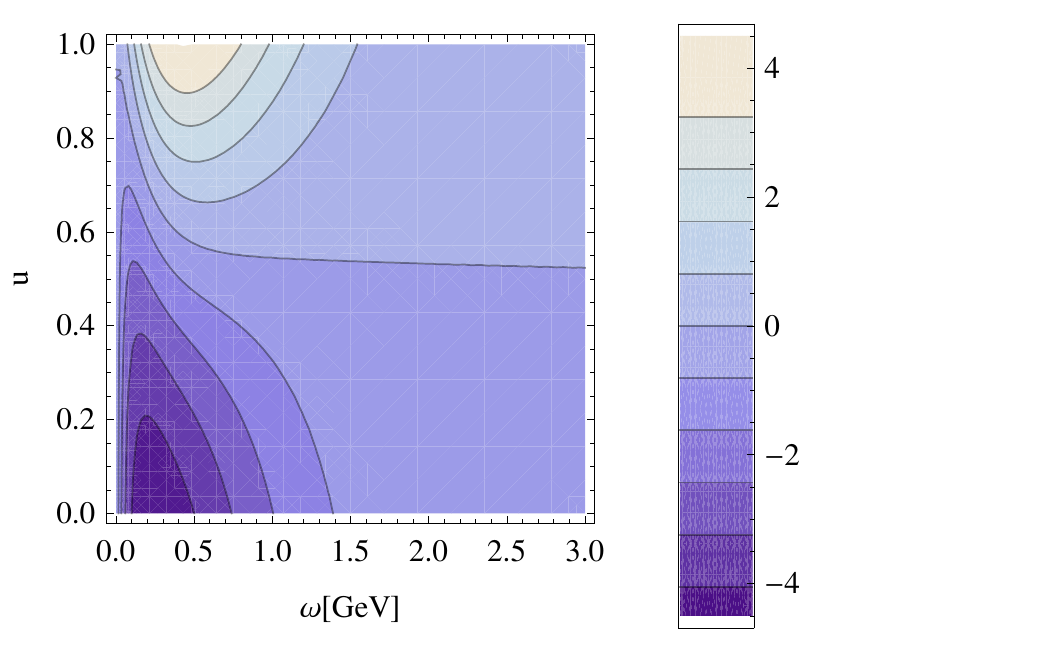}\hspace{-1.2cm}
\includegraphics[width=7cm]{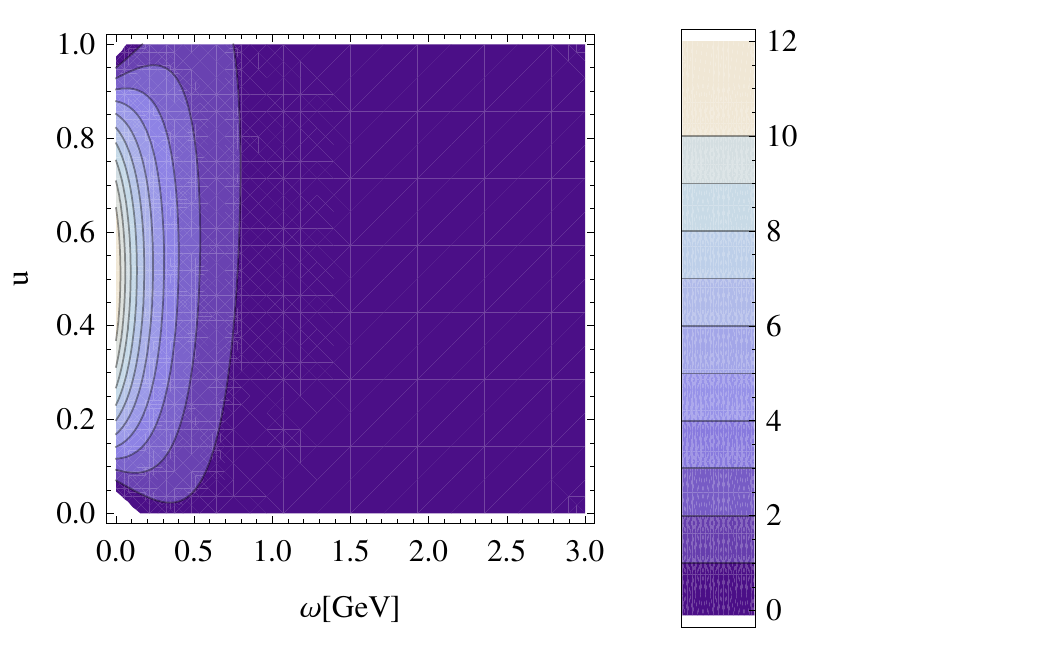}
\caption{ Model function $\psi(\omega,u)$ of the $b$-baryon $\Xi_b$. Shown are the twist 2 (upper left hand), symmetric twist 3  (upper right hand), antisymmetric twist 3  (lower left hand) and twist 4 (lower right hand) LCDAs.
All entires are evaluated at the energy scale  $\mu=1\; \tn{GeV}$.
\index{Model plots $b$-baryon LCDAs}
\label{xi3dplotexample}}
\end{figure}
%
%
With Tab.~\ref{momentsmodel} the model parameters are extracted 
from the moments and collected in Tab.~\ref{modelparam} in which  
the first moment is~$1$ by definition according 
to~\eqref{normalizationinlocallimit}, and the $\epsilon_i$ are 
required to be non-negative. The twist 2 model parameter dependence on $A$ is given in Fig.~\ref{tw2modpar} as example. 
To give an overview over the tables of model parameters, we show the
plots of the LCDAs in Fig~\ref{alltwist2onevariable} in the~$\omega$ 
and~$u$ distributions. The corresponding LCDAs for $\Xi_b$ ($\Xi'_b$) 
in the~($\omega$, $u$) plane are shown in Fig.~\ref{xi3dplotexample} 
as example.

\section{Conclusions}
\label{ch_can and out} 

As shown in Fig.~\ref{fig:SU(3)-multiplets} we are able to obtain 
the LCDAs for the entire ground-state multiplets of the bottom baryons
with one heavy quark, thereby generalizing the work by Ball, Braun and 
Gardi~\cite{Ball:2008fw}. We accounted for the mass-breaking effects due 
to the strange-quark mass and calculated the evolution of the LCDAs 
with the use of the renormalization group equations, obtained from 
the one-loop renormalization of the non-local light-cone operators 
given in Eqs.~\eqref{parallelcurrents} and~\eqref{transversalcurrents}. 
The resulting LCDAs are plotted in Figs.~\ref{alltwist2onevariable} and~\ref{xi3dplotexample} 
 in which the evolution from $\mu = 1$~GeV 
to $\mu = 2.5$~GeV and our error estimates are also shown.
We find that the $SU(3)_{\rm F}$-breaking effects are of order~10 percent. 
The sources for the $SU(3)_{\rm F}$ breaking are the strange-quark mass, 
the Borel parameter~$\tau$, the momentum cutoff~$s_0$ and the non-local 
condensates. 
The latter ones are not well determined in the case of the strange quark. 
However, the mass-breaking effects appear in the lower part of the $
\omega$-spectrum in the region of the constituent $s$-quark mass, 
as expected.
Even though the $SU(3)_{\rm F}$-breaking effects appear to be quite 
large at some momenta, they are within the conservative errors for 
the massless case, as is the evolution to the energy scale 
$\mu = 2.5 $~GeV. The errors are obtained by varying the parameter~$A$ 
in the full possible range $[0, 1]$, which defines the linear 
superposition of the two local interpolating currents, given 
in Eq.~\eqref{thelocalcurrentdefinition}. 
Reducing the variation of~$A$ to the less conservative range $[0.3, 0.7]$ 
results in a variation of the distribution amplitudes of roughly 
the same size as the evolution to $\mu = 2.5$~GeV.

\section*{Acknowledgements}

We thank V.~Braun, A.~Bharucha and Y.\,M.~Wang 
for fruitful discussions.
The work of W.~Wang is partly supported 
by the Alexander-von-Humboldt Foundation. 
A.\,Ya.\,P. thanks the Theory Group at DESY 
for their kind hospitality, where the major 
part of this paper was done.  
The work of A.\,Ya.\,P. was also performed 
within the research program approved 
by the Ministry of Higher Education of the 
Russian Federation for Yaroslavl State University
and was supported in part by the Russian Foundation
for Basic Research (project no. 11-02-00394-a) and 
by the Ministry of Education and Science of the 
Russian Federation in the framework of realization 
of the Federal Target Program ``Scientific 
and Pedagogic Personnel of the Innovation Russia'' 
for 2009-2012 (project no.~P-795).

\end{document}